\documentstyle[general_cite]{mn}
\begin{document}


\title[kiloparsec symmetric objects I. The candidates]{Flat-spectrum symmetric objects with $\sim$~1~kpc sizes\\ I. The candidates}
\author[Augusto et al.]{Pedro~Augusto,$^{1}$
J.~Ignacio Gonzalez-Serrano,$^{2}$  
 Ismael Perez-Fournon,$^3$ \and and Peter~N.~Wilkinson,$^4$\\
 $^{1}$Universidade da Madeira,
Centro de Ci\^encias Matem\'aticas, Caminho da Penteada, 9000-390 Funchal, Portugal\\
$^{2}$Instituto de F\'{\i}sica de Cantabria (CSIC-Universidad de Cantabria), 39005 Santander, Spain\\
$^{3}$Instituto de Astrof\'{\i}sica de Canarias, c/ Via L\'actea s/n, 38200 La Laguna, Tenerife, Spain\\
$^{4}$University of
Manchester, Jodrell Bank Observatory,
Macclesfield, Cheshire SK11 9DL, UK}


\bibliographystyle{mnras}
\maketitle

\begin{abstract}
In order to understand the origin and evolution of   radio galaxies, searches for the youngest such sources have been conducted. Compact-medium symmetric objects (CSO-MSOs) are thought to be the earliest stages of   radio sources, with possible ages of $\la10^3$~yrs for CSOs ($<1$~kpc in size) and 10$^4$--10$^5$~yrs for MSOs (1--15~kpc). From a literature selection in heterogeneous surveys,
we have established a sample of
37 confirmed CSOs.
In addition, we only found three confirmed flat-spectrum MSOs in the literature.
 The typical CSO resides on a $z\la0.5$ galaxy, has a flat radio spectrum ($\alpha_{thin}<0.5$; S$_{\nu}\propto \nu^{-\alpha}$), is $<0.3$~kpc in size, has an arm length ratio $\leq2$, and well-aligned ($\theta\leq20\degr$) opposite lobes with a flux density ratio $\leq10$. 
In order to populate the 0.3--1~kpc size range (large CSOs) and also in order to find more flat-spectrum  MSOs,
 we have built a sample of 157 radio sources with $\alpha_{1.40}^{4.85}<0.5$ that were resolved with the VLA-A 8.4~GHz. As first results, we have `rediscovered' nine of the known CSO/MSOs while identifying two new $\sim14$~kpc MSOs and two candidate CSO/MSOs (which only lack  redshifts for final classification). We were able to reject 61 of the remaining 144 objects from literature information alone. In the series of papers that starts with this one we plan to classify the remaining 83 CSO/MSO candidates (thanks to radio and optical observations) as well as characterize the physical properties of the (likely) many 0.3--15~kpc flat-spectrum CSO/MSOs to be found.
\end{abstract}

 

\section{Introduction}

\subsection{The evolution of   extragalactic radio sources}


 The origin and evolution of   extragalactic radio sources is
 one of the outstanding problems in Astronomy (e.g.\ \pcite{deVetal98a}) and has been a
 fundamental problem in the study of active galaxies and their nuclei
 (AGN). These come in a
 variety of sizes, from compact ($<$ 1 kpc) to very large ($>$ 1
 Mpc). This wide range of sizes has been interpreted as evidence for
 size evolution of the radio structure (e.g.\ \pcite{BlaRee74,Car85,Fanetal95,Reaetal96a},b).  
In the standard model of AGN, a central supermassive black hole,
 with $\sim10^6$--$10^9$ M$_{\odot}$ feeds on the material of the host galaxy to produce two   opposing radio
 emitting jets, thus
 creating a symmetric source that might only be disturbed by the
environment/speed of the jets, unless its source runs out of fuel. 
Mature radio galaxies fit into this picture and are mostly split up into
\scite{FanRil74} type~I and type~II radio galaxies (FRI and FRII). 
Up to 10$^5$ times smaller, compact-medium symmetric objects (CSO-MSO) might be their precursors (e.g. \pcite{Reaetal96a},b).

Traditionally, CSO/MSOs have always been considered high-power radio sources. However, low-power sources must be considered as well, if we really want to tell a story about the evolution of small (and young) radio sources all the way until becoming large radio galaxies (FRII {\em or} FRI) --- e.g. \scite{MarSpeKun03}. \scite{Fanetal95} already pointed out the hypothesis of MSOs evolving not into FRIIs but into FRIs and despite their bias towards high-power CSSs they concluded that, really, only the most powerful MSOs could be the precursors of FRIIs. Similar conclusions were reached by \scite{Reaetal96b}, while
\scite{Midetal04} go as far as proposing the radio structure of NGC7674 (a Seyfert galaxy) as the one of a (very weak) CSO. 
 \scite{Beg96} considered both hypothesis: lower power CSO/MSOs would evolve into FRIs while the high power ones would become FRIIs.
 In order to constrain models it is important to extend the radio power range \cite{Fanetal01}.

For example, the square-root decrease with size of the luminosity from CSOs to FRIIs proposed by \scite{Beg96} and \scite{Fanetal95}, using the border value of $1\times10^{25}$~W/Hz\footnote{The formal boundary from \scite{FanRil74} is at 178~MHz: $5.3\times10^{25}$~W/Hz with our cosmology. A typical radio galaxy ($\alpha_{0.178}^{1.4}=0.8$) has L$_{1.4}=1\times10^{25}$~W/Hz while a flat spectrum CSO/MSO ($\alpha_{0.178}^{1.4}=0.4$--0.5) has L$_{1.4}=2\times10^{25}$~W/Hz.}, implies that a $\sim1$~Mpc FRII evolved from a 10~pc CSO with a $>3\times10^{27}$~W/Hz power, through a 10~kpc MSO with $>1\times10^{26}$~W/Hz. This is why the total radio power of CSO/MSOs was assumed to be $\sim10^{26}$--$10^{27} h^{-2}$ W/Hz in earlier surveys (e.g. \pcite{PhiMut82,Reaetal96a,Fanetal95,Muretal99}). Recent surveys (e.g. \pcite{Kunetal05}) include much weaker sources.

`Hot-spots' in CSOs are so close ($<$1~kpc) to the nucleus that they might help towards the understanding of the central engines in AGN (\pcite{Reaetal96a},b). Furthermore,  
they
 are unique probes to
 the physics of the gas clouds of the broad line-emitting region ---
 \scite{Reaetal96a}. MSOs, being larger (1--15~kpc), are ideal to probe the ISM further away from the nucleus (including the clouds in the narrow line-emitting region (NLR) and extended NLR).
They might also be the middle link
  for the hypothetical evolution of CSOs into FRIIs or FRIs.

 Assuming the jet to travel at the speed
of light we get upper limits of $10^{3-4}$ years for $\sim1$~kpc
CSO/MSOs (large CSOs and small MSOs). Furthermore, kinematic measurements on ten CSOs
(\pcite{Giretal03,Ojhetal04}; e.g. \pcite{PolCon03})  give $v\simeq0.05$--$0.3h^{-1}$~c (hotspot advance speed) which, assuming constant speeds since source ignition (e.g.\ \pcite{Reaetal96b}), give ages of $\sim$300--2000~yrs. These are  consistent with sychrotron loss time scales  ($\sim1200$--5000~yrs; e.g.  \pcite{Reaetal96a,Giretal03}).
 CSOs evolve fast (c.f. FRIIs lobe advance speed  $\sim0.06$~c), explaining their `rarity':  only $\sim$10\% of radio sources with compact structure are CSOs \cite{Reaetal96a}, getting down to 1\% for the flat-spectrum ($\alpha_{1.40}^{4.85}<0.5$) largest CSOs and small MSOs \cite{papI}.

%
%


There are still two possibilities for the origin of CSO/MSOs, summarized in what follows. 

\begin{description}

\item[\bf Youth scenario:] The most popular view is that in which CSOs evolve into MSO/Compact Steep Spectrum Sources\footnote{The original definition is on \scite{PeaWal82}, with $\alpha_{2.7}^5\geq0.5$ (now as far as $\alpha_{0.325}^{1.4}$  --- e.g. \pcite{Tscetal03}), who also define almost half of their sample as ``compact'' and with ``steep'' spectra; in an historical perspective, up to this time {\it compact} $\Leftrightarrow$ {\it flat spectrum} and {\it extended} $\Leftrightarrow$ {\it steep spectrum}. \scite{PhiMut82}
demand an optically thin regime with $\alpha_{\nu_1}^{\nu_2}\geq0.5, \nu_1,\nu_2>1 \:$~GHz. We define $\alpha_{thin}$ from a full spectrum linear fit to the part that is optically thin for frequencies greater than a given peak; if there is no peak, it is inferred to lie at some still unobserved low frequency and the full spectrum is used.}
 (CSSs; $\alpha_{thin}>0.5$ with $S_{\nu}\propto \nu^{-\alpha}$)  which, in turn, evolve into FRIIs --- e.g. \scite{PhiMut82,Car85,Beg96,Reaetal96b,Kunetal02,PerMar02}. The intermediate $\sim1$~kpc stage should be a CSS in the case of self-similar expansion (lobes expand with growth) or a flat-spectrum MSO in case the expansion is non-self-similar (hot spots remain compact, if seen at all).
Maybe
less luminous CSOs evolve into FRIs via a Giga-Hertz Peaked Spectrum Source (GPS) stage
 \cite{ODe98,deVetal98b}.

%
%

\item[\bf Re-born scenario:] From an analytical model of the evolution of   double radio sources $<100$~kpc, \scite{Ale00} extended \scite{KaiAle97} model to $\sim$~kpc scales: a population of `short-lived' sources is predicted, where the jets are disrupted before reaching the $\sim1$~kpc core radius (King density profile) of the host galaxy. This could be interpreted in the context of ``re-birth''.
For example, \scite{Bauetal90} show the 47~pc CSO B0108+388 to have weak radio emission on tens of kpc scales; this might be an unrelated source or evidence for recurrent activity.

\end{description}


\subsection{Definitions}

\begin{figure*} 
\setlength{\unitlength}{1cm}
\begin{picture}(16,8)
  \put(0.5,0){\includegraphics{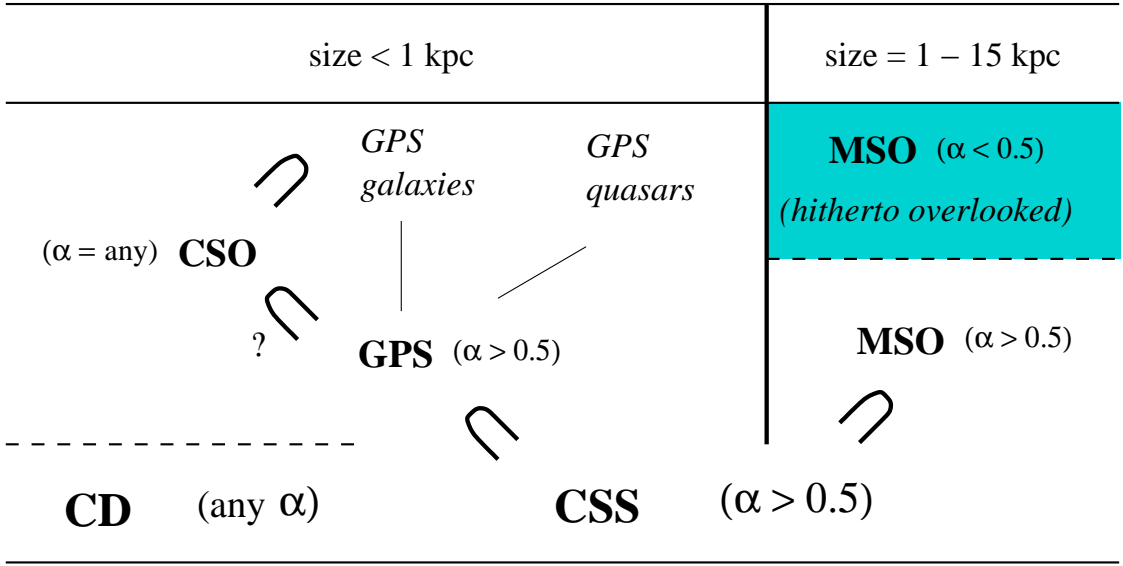}}
\end{picture}
\caption{The currently messy (and incomplete) situation in the definition of CSOs, MSOs, CSSs, GPSs and CDs. CDs and CSSs cover all ranges of sizes. The latter includes GPSs and MSOs, the first generally $<1$~kpc and the latter showing symmetric structure straddling a (putative or not) central nucleus.
 CSOs also appear with flat-spectrum ($\alpha<0.5$) and, in this sense, they cannot be paralleled to GPSs (even if only the ones identified with galaxies); in any case, there is no evidence for complete overlap of GPSs and CSOs. MSOs with a flat-spectrum (upper right corner) have been overlooked.
} 
\label{sumdef}
\end{figure*}

Over the last twenty years, a panoply of names have been used to classify $<15$~kpc-sized sources which might be the precursors of the much larger FRI/FRII radio galaxies.
Usually applied in the `young radio source' context, we have CSOs, MSOs, CSSs, GPSs, and, the oldest of all, compact doubles (CDs).
It is still disputed whether CSOs are included in the GPS class
 (e.g. \pcite{Sneetal99}, \pcite{ODe98}, \pcite{Maretal99} {\it vs.} \pcite{Staetal97}, \pcite{StaODeMur99}, \pcite{FasTay01}).
In Figure~\ref{sumdef} we summarize the current (confused) status and in Table~1
we propose a `non-grey zone' radio classification for all these sources, which can be used for the time being, at least operationally:
since CSO/MSOs  are a more homogenous class than GPSs are \cite{FasTay01}, we propose to split up the two main sets of `young sources' into the ones selected by morphology (CSO/MSOs) and the ones selected spectrally (CSS/GPSs). 
For CSO/MSOs, in particular, it should be made clear that it is not necessarily true that an edge brighened lobe is an hotspot. It might just be a knot in a longer jet. However, the likelihood that we get two of those opposed to each other and they are not hotspots is small. It is on this basis that CSO/MSOs with only two components are confirmed. When we come to triple (and more) component sources the hotspot/edge-brightened lobe definition relaxes: if we identify a central core component then we have a CSO/MSO structure (even if no obvious hotspots or edge-brightening is seen in any terminal lobe).

 Historically there has been a bias against steep-spectrum CSOs and flat-spectrum MSOs (e.g. the ``CSO-finding'' $\alpha<0.5$ Caltech-Jodrell Bank surveys (e.g. \pcite{Wiletal94}); the ``CSS ($\supset$MSO) finding'' $\alpha>0.5$ Bologna-Jodrell-Dwingeloo survey --- e.g. \pcite{Fanetal95}). 

\begin{table*}
\label{sumdefour}
\caption{Our proposed dual-view radio classification for CSOs, MSOs, CSSs and GPSs. CDs fall out of this picture since they can be all or none (when one-sided core-jet sources).
The scheme below must be viewed as operational only, since in the coming years more knowledge of the sources involved might suggest a different one. Note that we give strict morphological criteria to confirm the classification of CSO/MSOs; also that the latter are similar, apart from size (MSOs can be flat-spectrum radio sources).} 
\begin{center}
\begin{tabular}{cccc} \hline \hline
\multicolumn{2}{c}{\it Classification} & {\it Two components (lobes)} & {\it Three/more components} \\ \hline
 & CSO & edge-brightness clearly  & \\
morphological		& ($<$1~kpc) & seen in {\em both} & (one of the) central component(s)  \\
	  &  MSO & or kinematics show    & proved as the core    \\
		& (1--15~kpc) &		opposed motion&  		\\ \hline
  & & {\it Spectral turnover} & {\it GPS/CSS classification references} \\ \hline
& GPS	& 0.5--10~GHz & \protect \scite{Fanetal95} \\
spectral		&  ($<$1~kpc) &	 &  \protect \scite{deVetal97}  \\
($\alpha_{thin}>0.5$) & CSS	& $\la$0.1--0.5~GHz & \protect \scite{ODe98} \\
		&  ($<$15~kpc) &	 & \protect \scite{Tscetal03}  \\ \hline
\end{tabular}
\end{center}
\end{table*}

\subsubsection{CDs}

Compact double (CD) is the name from where all names come from: CSO/MSO/CSS/GPS. In fact, the classification CD remains, for example, when no core is seen in a candidate CSO/MSO.
\scite{PhiMut82} and \scite{Car85}
 ``compact symmetric (double) sources'' were defined with no central core and a ratio of flux density between the two lobes $\leq$1.5, $\alpha_{thin}\geq0.5$ (except 3C394 with 0.4), sizes $<$0.1~kpc, ages $10^3$--$10^4$~yrs, and $v\simeq0.2$~c (theoretical lobe advance speed).

\subsubsection{CSOs ($<1$~kpc)}

A CSO  \cite{Wiletal94,Conetal94} is a compact radio source with two outer edge-brightened lobes/hotspots or twin-jets plus a (possibly putative) central core. 
 Symmetry is essential so, operationally, the arm ratio should be $\leq10$, although the flux density ratio (between lobes) is not constrained (it is frequency-dependent).
CSOs have weak polarization and variability ($<10\%$ in a few years): some are so stable that they might be excellent VLBI flux density calibrators ---
 \scite{FasTay01}.

\subsubsection{CSSs ($<$15~kpc)}

This class of source, with a subgalactic size and a steep spectrum ($\alpha_{thin}>0.5$), has more pronounced lobe flux density ratios
 and/or arm ratios than CSOs \cite{Fanetal90,Fanetal95,Daletal95,Sanetal95,ODe98}.
When with a spectrum peak at $\nu>0.5$~GHz they are classified as GPSs (Section~1.2.4) while when symmetric (most --- e.g. \pcite{Reaetal96a,Fanetal95,Kunetal05}) they are called MSOs (Section~1.2.5).
They show low radio polarizations and little variability although up to an order of magnitude more variable than the most stable CSOs \cite{FasTay01}.


\subsubsection{GPSs ($<1$~kpc)}

In most properties, GPSs are similar to CSSs. The main difference is in the spectral peak (c.f. \protect Table~1; e.g. \pcite{Toretal01}): the canonical turnover frequency of GPSs is 1~GHz --- \scite{deVetal97}.
Also, many are
 highly variable (mostly identified with quasars --- \pcite{Toretal05}) jeopardizing their  usual classification when based only on sparse spectral data points (both in observing epochs and in frequency) --- 
 \scite{Staetal98} and \scite{Toretal01}.

\subsubsection{MSOs (1--15~kpc)}
So far regarded as steep-spectrum sources,
we here point out the existence of $\alpha<0.5$ flat-spectrum MSOs (c.f. Sections~2.3 and 4) as
 a hitherto not considered
 type of source (they fill the ``empty corner'' in Figure~1). \scite{papI} mention many candidates for such sources. Flat-spectrum MSOs could be the sources into which CSOs evolve when the expansion is non-self-similar \cite{DeY97,Tscetal00}. The statistics of MSOs are relevant in order to inspect which evolutionary scenario (non-self-similar vs. self-similar expansion) dominates.


\subsection{The optical hosts}

Not much is known at optical wavelengths about CSOs since only a few cases have been studied \cite{Tayetal96b}.
\scite{Reaetal96a} and  \scite{BonGarGur98} found that the hosts of five CSOs  are mostly $m_V\sim20$--22 elliptical galaxies (0.3--1~L$^{*}$) with strong narrow emission lines; the continuum is characteristic of an old stellar population. Detailed HST views 
 of three nearby ($z\la0.1$) CSOs \cite{Peretal01}
 confirm residence in normal ellipticals but with ten times more dust than radio elliptical galaxies.

A lot more is known in the optical about CSS/GPSs, which have similar redshift distributions and have as hosts $0.1\la z \la1$ regular giant elliptical galaxies (many interacting), like FRIIs do, a fact consistent with a GPS $\rightarrow$ CSS $\rightarrow$ FRII source evolution\cite{ODe98,deVetal00}.
GPS galaxies ($z\sim0.3$) show a CSO morphology while the quasars ($z\sim2$) do not \cite{ODeBauSta91,deVetal98b,Sneetal99,Staetal01}.

\subsection{This paper}

 The total number of confirmed CSOs is relatively small (37 --- Section~2; 25 have linear size information) for a two orders of magnitude size range (0.01--1~kpc). Worse, only three $\alpha<0.5$ flat-spectrum MSOs (1--15~kpc) and four `large CSOs' (0.3--1~kpc) are confirmed, so far.  The lack of `large CSOs' and flat-spectrum MSOs might be explained by a CSO `preferred' evolution into CSSs (e.g.\ Section~1.1; \pcite{papI}), but we need better statistics.

 The aim of the series of papers  which starts with this one is to find a fairly large number of $\alpha_{1.40}^{4.85}<0.5$ flat-spectrum CSO/MSOs with $\sim$kpc sizes (large CSOs and MSOs).
We start by establishing the current sample of confirmed CSOs as well as describing their overall properties (Section~2). In Section~3 we build up a 157-source sample
 out of which we expect a few tens to be confirmed as CSO/MSOs when our study  is complete. 
We conclude with a brief summary (Section~4).

Throughout the paper we use an $\Omega_m=0.3$, $\Omega_\Lambda=0.7$,  H$_0=75$~km/s/Mpc cosmology.

\section{Confirmed symmetric sources}

The literature abounds with examples of confirmed CSOs (summarized in Section~2.1) while MSOs are only abundant as CSSs, i.e., with a steep spectrum. Flat-spectrum MSOs are rare (Section~2.3).

\subsection{The sample of CSOs}

\begin{figure} 
\setlength{\unitlength}{1cm}
\begin{picture}(8,9)
  \put(10,0.5){\includegraphics{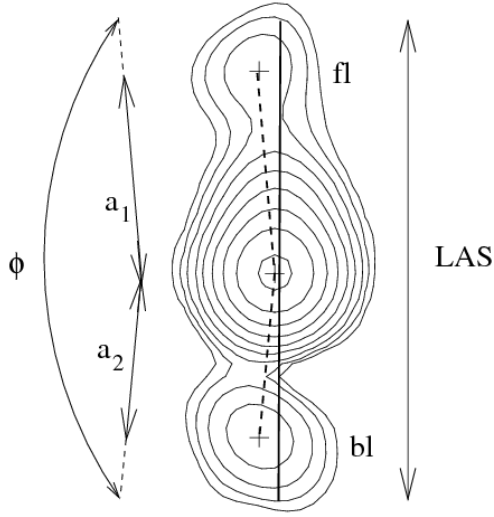}}
\end{picture}
\caption{In order to help understanding the parameters measured for each source in Columns~(10) and (13)--(16) of Table~2, we here present a case with an actually located core (without which, only Columns~(10), (13) and (14) might have values). Most maps in the literature are presented, like this one, with the lowest contour at three times the r.m.s. noise on the map (3$\sigma$). Then, in order to estimate its angular size (LAS), we measure the largest possible extent on the $6\sigma$ contour. Next, we locate the core and each lobe peak (marked with crosses) and immeditately identify the strongest lobe ($bl$) from the contours alone, deriving a peak flux density estimate. The same is made for the faintest lobe ($fl$) and the nucleus, whereby calculating the values in Columns~(13) and (14). Finally, joining by segments the crosses that correspond to the two lobes and the nucleus, we estimate $\phi$ (Column~(16)) and $R=\frac{a_1}{a_2}, a_1>a_2$ (Column~(15)).} 
\label{ex_par}
\end{figure}

In \protect Table~2 we present 
all currently known confirmed CSOs, proved as such from maps (or kinematics, in a few cases) in 
 our extensive literature search.
We were very rigorous in our classification, using the criteria of Table~1.
 All relevant maps/information have been compiled and carefully scrutinized before listing a given CSO as ``confirmed''  beyond any reasonable doubt. 
Everytime 
 the candidate had three or more components (even when some doubt remains about which of the central components really is the core), we required  a 
confirmed central core usually from, at least, two-frequency data.
If having only two components, they must be edge-brightened lobes: assumed is a putative central core (c.f. \pcite{TayReaPea96,TayVer97,BonGarGur98,Poletal99}). 
`Hotspots' are not necessarily required for sources with three (or more) components; all we need is emission on both sides of the core (even if jet-like). This is the usual way CSOs have been identified (see, for example, \pcite{Reaetal96a,Staetal97,PecTay00}).
 Since CSOs have sizes $<1$~kpc, we rejected all sources with size $>1$~kpc and
since, by definition, they are  symmetric sources (e.g. \pcite{Reaetal96a},b), we ruled out any with an arm ratio $R>10$.
 Figure~\ref{ex_par} defines and explains the calculation of the radio map parameters in Columns~(10) and (13)--(16) of  Table~2.

In five CSOs studied, \scite{TayReaPea96} find considerable flux density ratio asymmetries in the two opposing jets, possibly due to differences in density of the surrounding medium \cite{Staetal97}. Furthermore,
flux density ratios depend on frequency.
Hence, it seems more dangerous to place a limit on such ratio and we do not do it. We also do not constrain arm angles (column (16) of \protect Table~2) since, for example, we have $\phi=148\degr$ (misalignment $\theta$ is 32\degr) for a `classic' CSO (B2021+614) and only three CSOs in the Table are more misaligned (reaching a minimum of $\phi=134\degr$ for B1543+005, a \scite{PecTay00} CSO).

Comments on the sources marked with $^*$ in Column (1) of \protect Table~2  follow:
{\bf B0046+316:} This is a Sy2 galaxy; it possibly has a core-jet radio structure in a weird geometry \cite{Antetal02}. 
{\bf B0424+414, B0500+019, B0646+600, B0703+468, B0710+439:}  These sources are also classified as GPSs (e.g. \pcite{ODeBauSta91,Maretal99,Staetal01}).
{\bf B1934-638:} This is the archetype GPS (e.g. \pcite{Tzietal89}).



\subsection{Statistics}

Since all confirmed and candidate CSOs of Table~2 have been 
extracted from different samples in the
   literature with no other selection criteria except for morphology,
the statistical results must be taken
with
   caution since they might not be representative of the CSO class.
We  list 41 sources in \protect Table~2 out of which four (labelled ``CSO?'') still might be MSOs if their sizes turn out to be 1--15~kpc: we keep them in the table until we have enough data to finally classify them. This leaves us with 37 certain CSOs which we use in
 the statistical study that follows.

\begin{figure} 
\setlength{\unitlength}{1cm}
\begin{picture}(8,6.5)
  \put(-0.5,7.2){\includegraphics{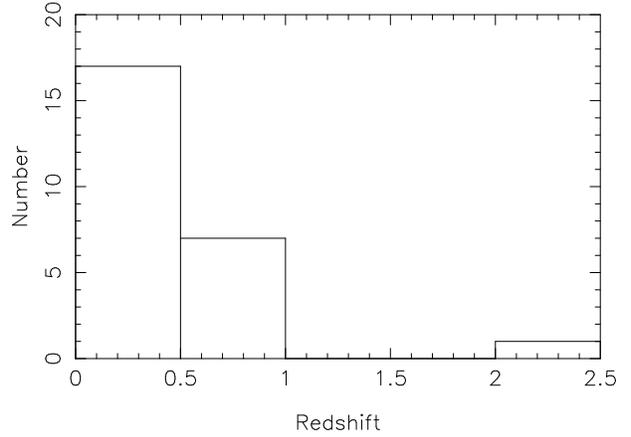}}
\end{picture}
\caption{The redshift distribution of 25 CSOs, with 68\% completeness.} 
\label{redshift_1}
\end{figure}

The optical information on the  27 CSOs that do have it (73\% completeness) shows that galaxies are clearly the typical host (23 or 85\%) while only four sources (15\%), at most, reside in quasars.
In \protect Figure~\ref{redshift_1} we present the redshift distribution of the sample (25 sources; 68\% complete). We clearly see a concentration towards low redshifts, with 17 (68\%) sources at $z<0.5$, implying a nearby galactic host population. In fact, except for one quasar, all CSOs  reside in $z<1$ hosts. The 25 CSO median is\footnote{The subscripts in the medians show the actual number of sources with values available for each calculation. We give the asymmetric error of the median at the 95\% confidence level.}  $z_{25}=0.36^{+0.16}_{-0.14}$. The quasar statistics (only three: they do not change the median at all) are still too poor to conclude that, like for GPSs, the hosts/redshifts imply two independent populations.

\begin{figure} 
\setlength{\unitlength}{1cm}
\begin{picture}(8,5)
  \put(-0.5,-4.2){\includegraphics{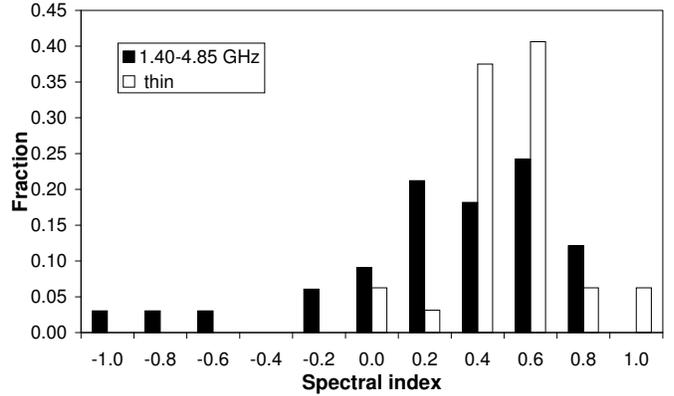}}
\end{picture}
\caption{The compared spectral indices ($\alpha_{1.40}^{4.85}$/$\alpha_{thin}$ --- see Table~2) distributions of 33/32 CSOs, with 89\%/86\% completeness.}  
\label{alpha_1}
\end{figure}

As regards to spectral indices, since we use a flat-spectrum sample defined from $\alpha_{1.40}^{4.85}<0.5$ (Section~3), we included this quantity for all CSOs in Table~2, in addition to $\alpha_{thin}$. Usually, to select CSOs or CSSs from weak samples, only a two frequency spectral index is used/available --- see e.g. \scite{Kunetal02}.
\protect Figure~\ref{alpha_1} shows the $\alpha_{1.40}^{4.85}$/$\alpha_{thin}$ distribution for 33/32 of the 37 CSOs of \protect Table~2 (89\%/86\% complete).
 Although CSOs, by definition, have
no spectral restrictions, it turns out that the majority have  a flat radio spectrum (23, or 70\% with $\alpha_{1.40}^{4.85}\leq0.5$; 18, or 56\% with $\alpha_{thin}\leq0.5$). The medians are $\alpha_{1.40}^{4.85}$ $ _{(33)}=0.3^{+0.2}_{-0.1}$ and $\alpha_{thin}$ $ _{(32)}=0.5\pm0.1$. There is a tendency for  $\alpha_{1.40}^{4.85}$  being flatter than $\alpha_{thin}$. In fact, defining $\Delta \alpha=\alpha_{1.40}^{4.85}-\alpha_{thin}$, we have $\Delta \alpha_{29}=-0.1\pm0.1$ (only 29 CSOs have both $\alpha_{1.40}^{4.85}$ and $\alpha_{thin}$ values available).

\begin{figure} 
\setlength{\unitlength}{1cm}
\begin{picture}(8,6.5)
  \put(-0.5,7.2){\includegraphics{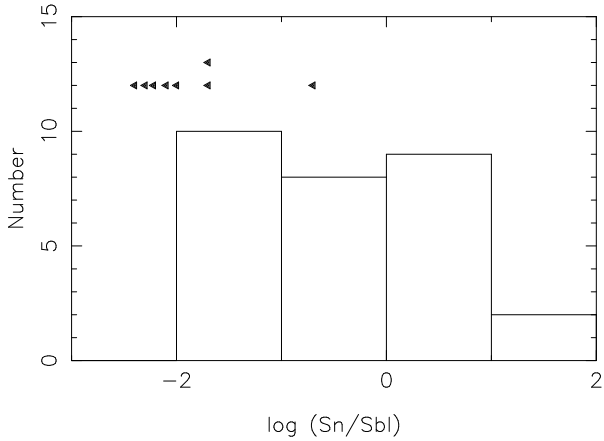}}
\end{picture}
\caption{The flux density ratio (nucleus over bright-lobe) distribution for 29 CSOs --- 78\% complete (the eight cases where no core is located are included as upper limits).} 
\label{snsbl_1}
\end{figure}

In \protect Figure~\ref{snsbl_1} we show the ratio in flux densities between core and brightest lobe for 29 CSOs (78\% complete). The remaining eight did not have a visible, properly located, core
but we still show the upper limits (estimated from the maps).
 It turns out that a bright nucleus (ratio $>1$ with respect to the bright lobe) is present in 11 (30\%) of the sources (including the eight sources with upper limits in the statistics), in one case about 40 times brighter. At the other end, five sources (with upper limits) have a nucleus more than 100 times weaker than the brightest lobe. The median is
  $S_n/S_{bl}$ $ _{(29)}=0.7^{+0.8}_{-0.6}$.

\begin{figure} 
\setlength{\unitlength}{1cm}
\begin{picture}(8,6.5)
  \put(-0.5,7.2){\includegraphics{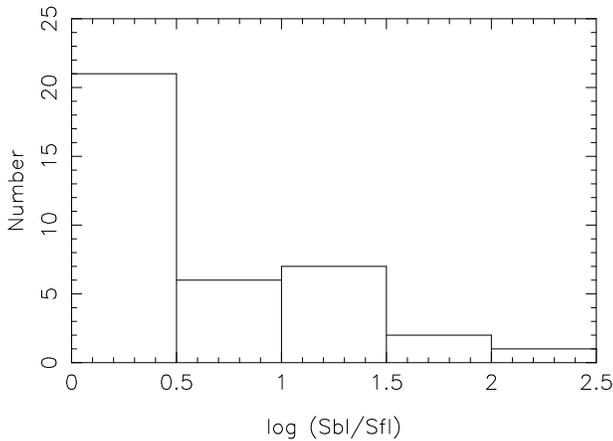}}
\end{picture}
\caption{The flux density ratio (bright-lobe over faint-lobe) distribution of all the 37 CSOs.} 
\label{sblsfl_1}
\end{figure}

We have studied the ratio in flux densities between 
the two opposed lobes for all the 37 CSOs (\protect Figure~\ref{sblsfl_1}): 27 (73\%) have ratios $<10$ (of which 21 (57\% of the total) of 1 to 3) but one CSO has it as large as 113: in total, ten (27\%) of the CSOs have $>10$ ratios. The median is
  $S_{bl}/S_{fl}$ $ _{(37)}=2.2^{+2.1}_{-0.2}$.

\begin{figure} 
\setlength{\unitlength}{1cm}
\begin{picture}(8,6.5)
  \put(-0.5,7.2){\includegraphics{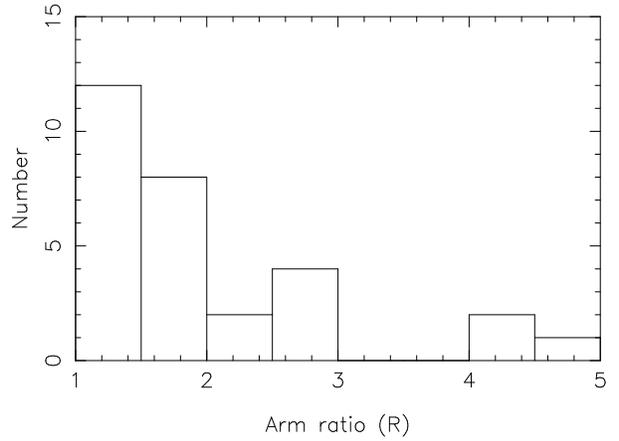}}
\end{picture}
\caption{The arm ratio ($R$) distribution of 29 CSOs, with 78\% completeness.}  
\label{arm_1}
\end{figure}

A test of symmetry, which was adopted as definition for CSO/MSOs, is the arm length ratio (\protect Figure~\ref{arm_1}). From the 29 CSOs (78\% complete) with data (core located, from where each arm is measured) we find that none has an arm ratio above 4.6 and that two-thirds (20) have it smaller than two. The spread is not large and we have for the median: $R_{29}=1.6^{+0.7}_{-0.4}$. Both in median and in distribution of arm ratios CSOs seem to lie somewhere between large FRII radio galaxies (symmetric) and CSS galaxies (asymmetric) --- e.g. \scite{Saietal03}.

\begin{figure} 
\setlength{\unitlength}{1cm}
\begin{picture}(8,6.5)
  \put(-0.5,7.2){\includegraphics{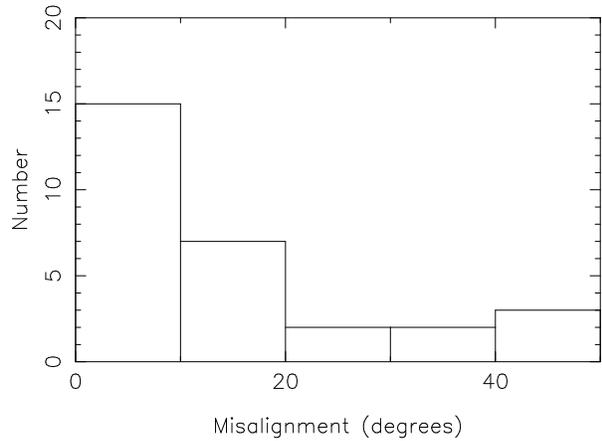}}
\end{picture}
\caption{The misalignment angle ($\theta$) distribution for 29  CSOs, 78\% completeness.}  
\label{align_1}
\end{figure}

Although not formally established, it is generally understood that a CSO/MSO should be fairly well aligned, similarly to FRI/FRIIs. The inter-arm angular ($\phi$)  distribution for 29 CSOs (78\% complete; they must have the core located in order to measure the angle) is plotted in Figure~\ref{align_1} but in the form of the
 misalignment angle ($\theta$), obtained by subtracting $\phi$ from 180\degr. 
We have the following medians:
 $\phi_{29}=171^{+5}_{-9}$ deg or $\theta_{29}=9^{+9}_{-5}$ deg.
We can then see that, as it was expected, CSOs are fairly well aligned sources, with $\theta\leq$20\degr for 22 (76\%) of them.

\begin{figure} 
\setlength{\unitlength}{1cm}
\begin{picture}(8,6.5)
  \put(-0.5,7.2){\includegraphics{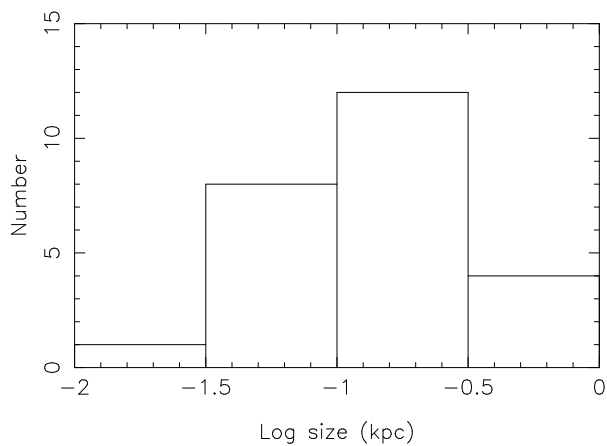}}
\end{picture}
\caption{The projected linear size ($l$) distribution of 25 CSOs (68\% complete).} 
\label{size_1}
\end{figure}

For the 25 CSOs (68\% complete) with measured redshifts, we plot, in Figure~\ref{size_1}, their projected linear size ($l$) distribution. The median is $l_{25}=0.14^{+0.07}_{-0.05}$~kpc. This indicates a lack of 0.3--1~kpc sources. In fact, from the Figure, we note that CSOs, in general, abound
at $<0.3$~kpc (21, 84\% of the total) but are scarce over the rest of the way up to 1~kpc.

\begin{figure} 
\setlength{\unitlength}{1cm}
\begin{picture}(8,6.5)
  \put(-0.5,7.2){\includegraphics{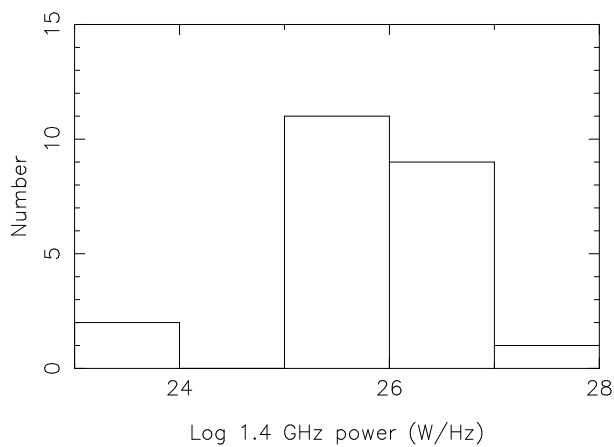}}
\end{picture}
\caption{The 1.4~GHz power distribution for 23  CSOs, 62\% completeness.}  
\label{power}
\end{figure}

In Figure~\ref{power} we show the 1.4~GHz power distribution for the 23 CSOs (62\% completeness) that have this information. The median [$\log L_{1.4}=25.9^{+0.4}_{-0.2}$ (W/Hz)] and the distribution clearly reflect the fact that the selection and classification of CSOs, so far, has implied high-luminosity sources, namely with $L_{1.4}>10^{25}$ (W/Hz). The bias is so strong that, using the formal definition at 178~MHz of the FRI/FRII border, with the help of the power extrapolation via $\alpha_{thin}$ and assuming a power decrease with the inverse square of size (e.g. \pcite{Beg96}), only one of the 23 CSOs in Table~2 with enough information will be powerful enough to become a 1~Mpc size FRII: 4C+32.44. If we relax their future to 100~kpc FRIIs, only two more will be added to the list: B0428+205 and 4C+62.22.
In any case,
we have searched for a correlation between size and 1.4~GHz radio power but found none. 

\subsection{Flat-spectrum MSOs}

We searched the literature and used the same criteria of Section~2.1 in order to identify $\alpha_{1.40}^{4.85}<0.5$ flat spectrum MSOs. Only three were found (sizes on 1--15~kpc):
4C+34.07, a $z=2.910$ QSO \cite{Wiletal98}; NGC3894, a $z=0.01075$ galaxy \cite{deVetal91}; B2151+174, a $z=0.23024$ cD galaxy \cite{Yeeetal96}. Their spectral and morphological radio properties (with references) are summarized in Table~3. 

\setcounter{table}{2}	
\begin{table*}
\label{MSOs}
\caption{\small The three  confirmed MSOs from the literature. The description for each column is as in Table~2 (columns~(10)--(18)). References for the radio maps on the three sources: \protect \scite{Daletal95,Beaetal02,Fometal00,Speetal89,Tayetal94,TayWroVer98,papI,Augetal04}.
}
\begin{center}
\begin{tabular}{lcclccccc} \hline \hline
\multicolumn{1}{c}{\bf (1)} & {\bf (2)} & {\bf (3)} & \multicolumn{1}{c}{\bf (4)} & {\bf (5)} & {\bf (6)} & {\bf (7)} & {\bf (8)} & {\bf (9)} \\
\multicolumn{1}{c}{Source} & $\alpha_{1.40}^{4.85}$  & $\alpha_{thin}$ & \multicolumn{1}{c}{LAS (\arcsec) [l (kpc)]}& S$_n$/S$_{bl}$ & S$_{bl}$/S$_{fl}$ & R & $\phi$&  L$_{1.4}$ \\ \hline
4C+34.07     & 0.41 & $0.3_s$& 1.85$^{1.7}$ (12.85) *& 11$^{1.7}$ & 6.7$^{1.7}$ & 1.2$^{1.7}$ & 174$^{\circ}$$^{1.7}$ & 26.8 \\
NGC3894     & $-0.33$ &  $-0.2_3^5$& 6.54 (1.34) &6.0&3.4 &1.0 & 180$^{\circ}$ &  22.9\\
B2151+174  & 0.13& $0.3_3$& 0.49 (1.58)& 15& 3.7&1.7 &166$^{\circ}$ & 25.2\\
 \hline
\end{tabular}
\end{center}
\end{table*}

\section{ Candidate $\sim$kpc flat-spectrum symmetric objects}

\subsection{The revised parent sample}

\scite{papI} have selected, from the $\sim4700$ sources of JVAS+CLASS1, a parent sample containing sources with
$\mid\!\! b^{II}\!\!\mid > 10^{\circ}$, $S_{8.4\, GHz}\geq100$ mJy and\footnote{ \scite{BonGarGur98} point out that at frequencies below $\sim1$~GHz, interstellar scintillation might induce extrinsic variability in extragalactic radio sources. Hence, our selection with $\alpha_{1.40}^{4.85}$ should be safe, as compared to other selections made with $\alpha_{0.3}^{high-\nu}$, likely more affected by such variability.}
$\alpha_{1.40}^{4.85}<0.5$ ($S_{\nu}\propto \nu^{-\alpha}$). However, their total of 1665 objects was short
by about 78 sources\footnote{Globally. There is a further complication since some sources that are in the revised sample were not in the old one (e.g. B0218+357) and vice-versa.}
because
 $\alpha_{1.40}^{4.85}$ was calculated from all kinds of catalogues. In order 
 to obtain
{\em same epoch} $\alpha_{1.40}^{4.85}$ 
 we got all values from \scite{WhiBec92} and \scite{GreCon91}, for the 1.40~GHz and 4.85~GHz frequencies, respectively\footnote{It would be more tempting to use the
 NVSS/GB6 combination (1.4/4.85~GHz; \pcite{Conetal98}/\pcite{Greetal96}), which would fill more blanks in Columns~(4)--(6) of \protect Tables~4 and 5. However,
given that the observation epochs are $\sim10$~yrs  apart, many such calculated spectral indices might not be trustworthy.
}. Sources without $\alpha_{1.40}^{4.85}$  information were also kept.

\begin{figure} 
\setlength{\unitlength}{1cm}
\begin{picture}(8,6.5)
  \put(-0.5,7.2){\includegraphics{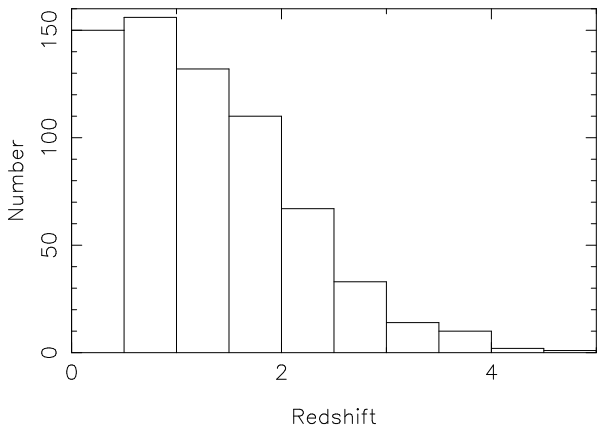}}
\end{picture}
\caption{ The redshift distribution of the 675 sources of the 1743-source parent sample which have such information (39\% completeness).} 
\label{zdist}
\end{figure}

\begin{figure} 
\setlength{\unitlength}{1cm}
\begin{picture}(8,6.5)
  \put(-0.5,7.2){\includegraphics{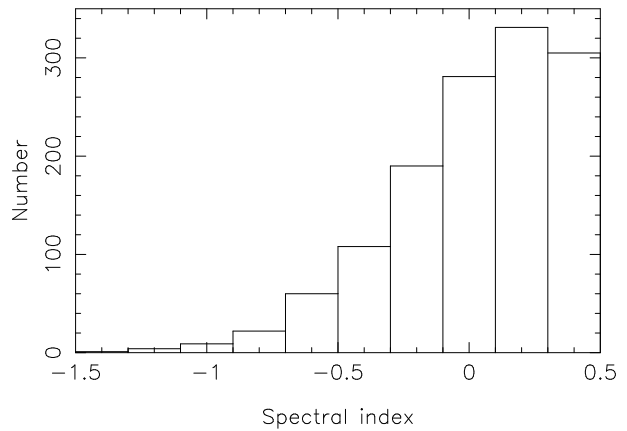}}
\end{picture}
\caption{ The 1.40--4.85~GHz spectral index distribution of the 1311 sources of the 1743-source parent sample which have such data (75\% completeness).} 
\label{a_dist}
\end{figure}

The revised parent sample now contains 1743 sources (\protect Table~4), whose 
redshift and spectral index distributions are presented in Figures~\ref{zdist} and~\ref{a_dist}, respectively. 
We note that these distributions use the full sample rather than just a representative subsample (c.f. \pcite{papI}).
A full discussion on the implications of the revision of the parent sample is made in Appendix~A.

\begin{table*}
\label{parent}
\caption{The parent sample of 1743 flat-spectrum ($\alpha_{1.40}^{4.85}<0.5$) radio sources (extract only --- the complete version can be found at CDS, ftp://cdsarc.u-strasbg.fr). Description of each column: {\bf (1):}  The source name (J2000.0); {\bf (2,3):} position (J2000.0);   {\bf (4):} 1.40~GHz flux density from \protect \scite{WhiBec92}  --- generally; $<110$~mJy conservative upper limits are placed on some sources which were covered in the sky survey but were not detected down to the $\sim100$~mJy treshold; other limits are for sources not covered in the survey and observed with NVSS --- \protect \pcite{Conetal98} (total flux density of all detected components within a 10\arcmin radius); {\bf (5):} 4.85~GHz flux density from \protect \scite{GreCon91};   {\bf (6):} spectral index, calculated from Columns (4) and (5) using the convention S$_{\nu}\propto\nu^{-\alpha}$; {\bf (7):} redshift; {\bf (8):} reference for the redshift; {\bf (9):} note/comment.} 
\begin{center}
\begin{tabular}{ccccccccc} \hline \hline
{\bf (1)} &  {\bf (2)} & {\bf (3)} & {\bf (4)} & {\bf (5)} &  {\bf (6)} &  {\bf (7)} &  {\bf (8)} &  {\bf (9)} \\
Name & RA (2000) & Dec (2000) & S$_{1.40}$ & S$_{4.85}$ & $\alpha_{1.40}^{4.85}$ & z & Ref. & Note \\
 &  &  & (mJy) & (mJy) & & & &  \\ \hline
J0457+067	&	04 57 07.7102	&	06 45 07.275	&	$<571$	&	435	&	$<0.22$	&	0.405	&	45	& \\
J0458+201	&	04 58 29.8726	&	20 11 35.997	&	170	&	163	&	0.03	&		&		& \\
J0459+024	&	04 59 52.0509	&	02 29 31.176	&	1752	&	1689	&	0.03	&	2.384	&	7	& \\
J0501+139	&	05 01 45.2706	&	13 56 07.218	&	235	&	468	&	-0.55	&		&		& \\
J0501+714	&	05 01 45.7829	&	71 28 33.977	&	$<110$	&	148	&	$<-0.23$	&		&		& \\
J0502+061	&	05 02 15.4466	&	06 09 07.507	&	1016	&	929	&	0.07	&	1.106	&	45	& \\
J0502+136	&	05 02 33.2194	&	13 38 10.949	&	581	&	504	&	0.11	&		&		& \\
J0503+020	&	05 03 21.1972	&	02 03 04.674	&	2118	&	1888	&	0.09	&	0.58457	&	21	& \\
J0503+660	&	05 03 56.4447	&	66 00 31.503	&	$<110$	&	158	&	$<-0.29$	&		&		& \\
J0505+049	&	05 05 23.1850	&	04 59 42.723	&	659	&	964	&	-0.31	&	0.954	&	45	& \\
J0505+641	&	05 05 40.9360	&	64 06 26.316	&	356	&	214	&	0.41	&		&		&	\\
J0508+845	&	05 08 42.3648	&	84 32 04.543     & & & & & &                                A 	\\
	 \hline
\end{tabular}
\end{center}
\end{table*}

\subsection{The 157-source sample }

Although small-size (VLBI scale: 1--300~pc) CSOs had dedicated searches/surveys in order to find them (e.g. \pcite{PecTay00}), bringing the current number of  confirmed cases to 37 (c.f. Table~2 and Figure~9), the  problem is that large-scale CSOs and $ \alpha<0.5$ flat-spectrum MSOs (0.3--15~kpc size range; see also Table~3) have only seven confirmed cases. It was vital first, no doubt, to establish CSOs as  new, worth of studying, objects and the VLBI efforts had the ideal impact, showing (many of) them as young sources. We believe that the time has come to start populating the 0.3--1~kpc size-range with CSOs, if we really want to learn about the full story of the small/young 1--300~pc CSOs evolution all the way into FRIIs or FRIs.
In this respect, more flat-spectrum MSOs are also needed.

 \scite{papI} presented 23 CSO/MSO  candidates included in a 55-source sample
selected from the parent sample by showing a greater than 25\% decrease in their 8.4~GHz VLA-A visibilities (usually corresponding to strong radio features with $<$7:1 flux density ratio and $\ga0\farcs1$ apart; at $z>0.2$, the projected linear size is $\ga0.3$~kpc). 
However, the 55-source sample was
 biased towards finding  gravitational lenses:
many sources were excluded from their final sample using a further surface brightness criterion in that sources with a bright and compact component plus other fainter and resolved components would be rejected (as well as any sources with dominant components $>0\farcs3$ apart) --- the full details are in \scite{Aug96}. 

Since we now do not apply to the parent sample any of those extra criteria and only use the visibility criterion,
we end up with a sample that includes 157 objects (including the 55 sources of \pcite{papI}) --- Table~5, some of which are very extended objects. We do not claim that this sub-sample of 157 CSO/MSO candidates is complete since neither is JVAS for the reasons presented, e.g, in \scite{papI} and \scite{Patetal92a}. 

Using the same classification criteria as for building Table~2 (mentioned in Table~1), 
 a literature search ruled out 61 sources as CSO/MSO candidates, also ``rediscovering'' nine of the CSO/MSOs in Table~2/Section~2.3 (including ``CSO?'' cases).
 We have discovered two new $\sim14$~kpc MSOs (J0751+826, J1454+299) and two CSO/MSOs (4C+66.09, J2055+287; a redshift is needed for each, for final classification; the latter case can be an MSO only if  its redshift is $<0.09$).

An obvious test to our criteria would be to check how many of the northern hemisphere confirmed flat-spectrum ($\alpha_{1.40}^{4.85}<0.5$) CSOs in Table~2 were not selected by us from the parent sample, and why. To identify them it is easy, since they are the ones that are  both in Tables~2 and~4  and not in Table~5: they total to 13, of which 12 have sizes $\leq40$~mas, hence they would never have been selected by our criteria. 
Recalling that we typically can only identify sources that have, at least, two components $\ga0\farcs1$ apart and with $<7:1$ flux density ratio at 8.4 GHz, 
as regards to the only remaining 
(B1413+135), although with a global size of $\sim0\farcs18$, it has a very faint lobe (S$_n/$S$_{fl}\simeq35$) while the bright lobe (S$_{n}/$S$_{bl}\simeq1.6$) is too close to the nucleus ($\sim$35~mas).  So, it could not have been selected.

In what follows we describe in detail the sources that refer to the text in Table~5:

\begin{description}
\item[\bf J0013+778:] This is a bright core large symmetric object (LSO), which has detailed information with 1.6~GHz VLBI \cite{Poletal95} and 1.4~GHz VLA  \cite{Xuetal95}. We locate its core thanks to the JVAS map, since at 8.4~GHz with the VLA-A the middle VLBI component has the most inverted spectrum of all while the northeast component also shows a more modest inverted spectrum; the one of the southwest component is very steep. The overall size is about 8\arcsec\/ in a north-south direction (thanks to further, weaker VLA 1.4~GHz components) which, at its redshift of 0.326, gives it $\sim32$~kpc.
\item[\bf 4C+36.01:] This is a radio galaxy with an extended halo, giving it an overall size of $\sim40$~kpc, as can be seen in a VLA 1.4~GHz map \cite{Tayetal96}. 
\item[\bf J0123+307:] This source is a good example on why JVAS is not a complete sample. As explained (and imaged) in detail in \scite{Aug96} this is, in fact, a VLBA point source (500:1 map) that had its position in error by an amount sufficient to cause bandwidth smearing and confuse our visibility selection criterion (see also \pcite{papI,Patetal92a}).
\item[\bf J0259+426:]   As for the previous source, this was also a JVAS failure and the selection was made erroneously. Although not quite a VLBI point source (it is a triple source $\sim15$~mas in size --- \pcite{Henetal95}) it should never have been selected.
\item[\bf  3C108:] This is a triple source with a candidate core at the centre from a MERLIN 1.7~GHz map in \scite{Saietal90}. In JVAS, the 8.4~GHz VLA-A map confirms the central component as a core ($\alpha_{1.7}^{8.4}\simeq0.2$) while the source redshift of 1.215 implies that its 5\farcs88 angular size means a linear projected size of $\sim$40~kpc. Hence, this source is rejected.
\item[\bf J0654+427:]   \scite{Bonetal01} give two VLBI maps at different resolutions (and frequencies: 1.6 and 4.9~GHz) that leave no room for doubt that the structure is that of a core-jet source rather than a CSO/MSO: the brightest component in both images is the core since it has $\alpha_{1.6}^{4.9}=0.0$ (using peak brightnesses), likely becoming inverted if model fitting is applied.
\item[\bf J0656+321:] Yet another source that is an example of why JVAS is not a complete sample. As explained (and imaged) in detail in \scite{Aug96} this is, in fact, a MERLIN point source (670:1 map) that had its position in error by an amount sufficient to cause significant bandwidth smearing (see also \pcite{papI,Patetal92a}).
\item[\bf J0751+826:]   Also presenting VLBI compact structure \cite{Poletal95,Xuetal95}, with an easily identified core, the large scale structure of this source, easily seen in the VLA-A 8.4~GHz map of JVAS, looks like a $\sim2\arcsec$ wide-angle tail. Its 1.991 redshift implies a global size of $\sim14$~kpc, just at the border of still classifying it as an MSO.
\item[\bf J0815+019:]  This source is in the appendix of \scite{papI}, discarded by them from the 55-source sample (due to an erroneous spectral index evaluation --- see Section~3.1). It is now recovered into the 157-source sample. \scite{papI} presented a MERLIN 5~GHz map of the source. We must locate the core in future multi-frequency follow ups.
\item[\bf J0817+324:] As explained (and imaged) in detail in \scite{Aug96} this is, in fact, a MERLIN point source (500:1 map) that had its position in error by an amount sufficient to cause significant bandwidth smearing (see also \pcite{papI,Patetal92a}).
\item[\bf J0837+584:]   All evidence seems to point to a core-jet source. In addition to the JVAS map and visibility, hinting at a strong unresolved component plus a very weak and distant ($\sim0\farcs6$ away) blob, the two frequency 1.6 and 5~GHz VLBI maps of \scite{Poletal95} and \scite{Xuetal95} locate the nucleus as the westernmost component, with $\alpha_{1.6}^{5}=0.15$ as opposed to the $\alpha_{1.6}^{5}=1.4$ value of the other strong component $\sim8$~mas
away.
\item[\bf J0855+578:] This source was observed with the VLBA at 5 GHz by \scite{Tayetal05}. It has one of the lobes edge-brightened but the other is not so convincing. There is no core detected. We must  find a core with higher frequency observations, or a more convincing structure to pass our strict criterion for a CSO confirmation; \scite{Tayetal05} have not managed to detect this (weak) source at 15~GHz.
\item[\bf 4C+66.09:]  This  source is in the appendix of \scite{papI}, discarded by them from the 55-source sample. It is now recovered into the 157-source sample. \scite{papI} presented a MERLIN 5~GHz map of the source which leaves no room for doubt that this source is either a CSO or an MSO, depending on its unknown redshift: edge-brigthening is seen in both lobes, although no core is detected. VLBA observations have been conducted for  this source at 1.7, 4.8 and 15~GHz \cite{Rosetal05} confirming the classification and finding hotspots at both ends, although the nucleus still remains undetected. 
\item[\bf 4C+55.17:] The multi-frequency maps of \scite{Reietal95} and the source redshift of 0.909 imply a large size ($\sim53$~kpc), although the source is symmetric indeed (an LSO).   
\item[\bf J1015+674:] \scite{Aug96} shows it as a MERLIN point source (300:1 map) that had its position in error by an amount sufficient to cause significant bandwidth smearing (see also \pcite{papI,Patetal92a}).
\item[\bf J1041+525:] This is a well-studied large scale quasar ($\sim150$~kpc in size), easily seen also in VLBI scales (e.g. \pcite{Henetal95,Tayetal96}).
\item[\bf J1058+198:] With a global size of $\sim420$~kpc (62\arcsec at a redshift of 1.11), this is a very large source, possibly a radio galaxy (e.g. \pcite{Hooetal92}).
\item[\bf J1110+440:] \scite{Aug96} shows it as a core-jet source (a very compact and strong nucleus and an extended, 50 times weaker, jet). 
\item[\bf J1306+801:] This is a very large triple source ($\sim$110~kpc --- \pcite{Tayetal96}), possibly an LSO, since the core appears to be located in the middle component (from multi-frequency data).
\item[\bf J1324+477:] \scite{Aug96} shows it as a point source, in a 200:1 MERLIN~5~GHz map, so this source was erroneously selected due to bandwidth smearing. 
\item[\bf J1424+229:] This is a well known arcsecond-scale gravitationally lensed multiple-image system (e.g. \pcite{Patetal92b}).
\item[\bf J1440+383:] This source is in the appendix of \scite{papI}, discarded by them from the 55-source sample and now recovered into the 157-source sample. \scite{papI} presented a MERLIN 5~GHz map of this double source for which its 8\arcsec\/ separation translates into $\sim50$~kpc at the source redshift of 1.775.
\item[\bf J1454+299:]  This  source is in the appendix of \scite{papI}, discarded by them from the 55-source sample and now recovered into the 157-source sample. \scite{papI} presented a MERLIN 5~GHz map of the source which leaves no room for doubt that this source is an MSO, given the edge-brigthening in both lobes and the presence of a central compact ``core''; the overall size of $\sim$2\farcs5 corresponds to $\sim14$~kpc at the source redshift of 0.58.
\item[\bf J1504+689:] \scite{Laretal01} show this source as a large-scale giant radio QSO, with a size of $1.16$~Mpc.
\item[\bf J1526+099:] A confirmed LSO (from the VLA-A maps of JVAS at 8.4~GHz and of \scite{HinUlvOwe83} at 1.4~GHz) which, given its redshift of	1.358  and from its angular size of $\sim15\farcs5$, has a global size of $\sim110$~kpc.
\item[\bf Arp220:] This is a very well known radio galaxy with ultraluminosity at IR wavelengths, presenting a double radio/IR nucleus \cite{Nor88,Graetal90} and also maser emission. Too many observations at all wavelengths exist for this source to mention here, so just as essential examples we cite \scite{Emeetal84,Soietal84,Noretal85,Shaetal94,Hecetal96,Scoetal98,Cleetal02}. It is not a CSO/MSO since it is thought that most of its radio emission comes from strong starburst activity (e.g. \pcite{Rovetal03}). 
\item[\bf 4C+49.26:]   As already pointed out by \scite{papI} this source is an LSO with a 6\arcsec\/ size which, at its redshift of 0.7, makes it $\sim36$~kpc in total.
\item[\bf J1607+158:] This is a core-jet source, from VLBI \cite{Beaetal02} up to 8.4~GHz VLA-A scales (JVAS). 
\item[\bf 4C+12.59:]   From several multi-frequency maps \cite{Saietal90,LonBarMil93,Daletal98} it is still not clear whether this source is a core-jet or an LSO. From our point of view this is irrelevant, since its angular size of 3\farcs3 and redshift of 1.795 make it $\sim24$~kpc in size.
\item[\bf J1715+217:] A recent VLBA map on this source (Gurvits et al. 2006, in prep.) shows it as a core with a  jet containing a strong feature about $\sim60$~mas from the core. The VLA-A 8.4~GHz visibilities, however, suggest larger scale structure as well. Future MERLIN 5~GHz observations should find it. 
\item[\bf J1749+431:]   All extant multi-frequency maps \cite{Henetal95,Tayetal96,Beaetal02} strongly suggest that this source has a core-jet structure.
\item[\bf J1753+093:] All evidence seems to identify this radio source with a galactic star \cite{ThoDjoCar90}. 
\item[\bf NGC6521:] \scite{ConCotBro02} find it likely that this source has a core plus two lobes on each side, with an overall size of 5\arcmin, giving it a size of $\sim150$~kpc at its 0.027462 redshift. 
\item[\bf NGC6572:] This source is a galactic planetary nebula (e.g. \pcite{ConKap98}).   
\item[\bf J2055+287:]  This source is in the appendix of \scite{papI}, discarded by them from the 55-source sample. It is now recovered into the 157-source sample. \scite{papI} presented a VLA 1.4~GHz map of the source which shows it with a clear structure containing edge-brightened lobes. If it lies at a redshift closer than  0.09  it still can be classified as an MSO (size $<15$~kpc).
\item[\bf J2234+361:]   This source is in the appendix of \scite{papI}, discarded by them from the 55-source sample and now recovered into the 157-source sample. \scite{papI} presented both MERLIN and VLBA 5~GHz maps of this source after which there is no doubt to classify its structure as a core-jet.
\end{description}

\begin{table*}
\label{157_sample}
\caption{The sample of 157 flat-spectrum ($\alpha_{1.40}^{4.85}<0.5$) radio sources. The columns are as in \protect Table~4, except that; i) column~(7) only exists here (its caption is as in column~(12) of Table~2 with an additional subscript: f --- flattening of the spectrum at high-$\nu$); ii) in column~(10) we give comments on the status of the candidate: CSO/MSO (?) --- (not yet) confirmed CSO/MSO --- see text (t) or \protect Table~2/Section~2.3 (S) for more details; OUT --- ruled out, see why in text or in \protect \scite{papI} for the sources marked with a ``55'' superscript; OBS --- data exist (need processing/interpreting); ? --- the cases that will need observations in the future for the first step in the classification of their structure; MERLIN/VLBI --- sources that have adequate structure but lack multi-frequency observations to confirm core location (see text or \protect \scite{papI} for the sources marked with a ``55'' superscript). For this table, also, we give here the redshift references (Column (9)) with the same code numbers as in \protect Table~4: 3-- \protect \scite{WhiKinBec93}; 7-- \protect \scite{HewBur89}; 9-- \protect \scite{WilWil76}; 10-- \protect \scite{Xuetal94}; 13-- \protect \scite{StiKuh93a}; 17-- \protect \scite{VerTay95}; 18-- \protect \scite{GonVerVer98}; 20-- \protect \scite{HewFolCha95}; 22-- \protect \scite{Hooetal96}; 23-- \protect \scite{Henetal97}; 25-- \protect \scite{Maretal96}; 37-- \protect \scite{MilOwe01}; 41-- Parkes Catalogue (1990), Australia Telescope National Facility, Wright \& Otrupcek, (Eds); 45-- \protect \scite{Drietal97}; 62-- \protect \scite{Alletal88}; 63-- \protect \scite{BurCro79}; 64-- \protect \scite{StiKuh96}; 65-- \protect \scite{Pucetal92}; 66-- \protect \scite{Stietal96}; 67-- \protect \scite{LeBetal91}; 68-- \protect \scite{Baletal73}; 69-- \protect \scite{Ungetal86}; 70-- \protect \scite{Sar73}; 71-- \protect \scite{Veretal96}; 73-- \protect \scite{FalKocMun98}; 81-- \protect \scite{HewBur91}; 86-- \protect \scite{deVetal91}; 103-- \protect \scite{Patetal92b}; 104-- \protect \scite{HooMcM98}; 108-- \protect \scite{Wegetal99}; 113-- \protect \scite{DonGhi95}; 141-- Sloan Digital Sky Survey ({\tt www.sdss.org}); 142-- \protect \scite{CohLawBla03}; 143-- \protect \scite{GorKonMin03}; 144-- \protect \scite{Magetal04}; 145-- \protect \scite{SowRomMic03}.}
\begin{center}
\begin{tabular}{cccccccccc} \hline \hline
{\bf (1)} &  {\bf (2)} & {\bf (3)} & {\bf (4)} & {\bf (5)} &  {\bf (6)} &  {\bf (7)} &  {\bf (8)} &  {\bf (9)} &  {\bf (10)} \\
Name & RA (2000) & Dec (2000) & S$_{1.40}$ & S$_{4.85}$ & $\alpha_{1.40}^{4.85}$ & $\alpha_{thin}$ & z & Ref. & Notes \\
 &  &  & (mJy) & (mJy) & & & &  \\ \hline
J0000+393	&00 00 41.5259	&39 18 04.172&	220	&138	&0.38& 0.4$_3$ & & & ? \\		
J0009+400	&00 09 04.1750	&40 01 46.724&	569	&333	&0.43&0.5 & 	1.83&	7& OBS\\
J0013+778	&00 13 11.6992	&77 48 46.620&	2203	&     &	&  0.5$_3^{0.1}$ &    	0.326	&10& OUT\\
J0020+430	&00 20 49.9798	&43 04 38.329	&363	&253	&0.29& 0.5 & & & ? \\		
J0026+351	&00 26 41.7238	&35 08 42.285	&819	&453	&0.48	&0.4$_3$ & 0.333	&13& ? \\
J0036+318	&00 36 48.1263	&31 51 14.532	&256	&148	&0.44& 0.6$_3$ & & & ?\\		
4C+36.01	&00 37 46.1437	&36 59 10.928	&879	&482	&0.48	&0.6$_s$ & 0.366	&17& OUT\\
4C+12.05	&00 38 18.0173	&12 27 31.252	&1002	&670	&0.32	&0.7$_f$ & 1.395	&18& ?\\
J0048+319	&00 48 47.1438	&31 57 25.094	&270	&254	&0.05	&0$_v$ & 0.015	&69& CSO (S) \\
J0112+203	&01 12 10.1864	&20 20 21.789	&407	&247	&0.40	&0.6$_3$ & 0.746	&7& ?\\
J0115+521	&01 15 56.8741	&52 09 13.034	&328	&206	&0.37 & 0.7 & & & MERLIN$^{55}$ \\		
J0119+321	&01 19 34.9991	&32 10 50.013	&2826	&1571	&0.47	&0.4$^{0.4}$ & 0.0592 & 70& CSO (S)\\	
J0123+307	&01 23 02.2783	&30 44 06.847	&126	&184	&$-$0.30& & & & OUT\\		
J0129+147	&01 29 55.3484	&14 46 47.843	&706	&536	&0.22 & 0.6$_f$ & 1.62985&141 & OUT$^{55}$ \\		
J0138+293	&01 38 35.3234	&29 22 04.544	&324	&184	&0.46 & 0.5$_3$ & & & ? \\		
J0209+724	&02 09 51.7921	&72 29 26.669	&842	&560	&0.33	&0.4& 0.895&	71& MERLIN$^{55}$ \\
J0221+359	&02 21 05.4702	&35 56 13.722     &	1456	&1498	&$-$0.02&0.3 & 	0.944&	142& OUT$^{55}$ \\
J0227+190	&02 27 53.3347	&19 01 14.082	&292	&160	&0.48& 0.5$_3$ & & & MERLIN/VLBI$^{55}$ \\		
J0237+437	&02 37 01.2149	&43 42 04.191	&431	&246	&0.45& 0.5$_3^{0.4}$ & & & CSO (S)\\		
J0255+043	&02 55 55.4349	&04 19 40.588	&435	&278	&0.36& 0.5$_3$ & & & ?\\		
J0259+426	&02 59 37.6753	&42 35 49.908	&616	&366	&0.42	&0.2 & 0.867	&17& OUT\\
J0308+699	&03 08 27.8276	&69 55 58.900	&228	&205	&0.09&0.3$_f$ &  & & OBS\\		
J0348+087	&03 48 10.4178	&08 42 08.873	&248	&192	&0.21& 0.5$_3$ & & & OUT$^{55}$  \\		
J0354+801	&03 54 46.1258	&80 09 28.816	&818 & & &0.4 &  & &? \\	     	     		
J0355+391	&03 55 16.5912	&39 09 09.824	&160	&191	&$-$0.14&0.4& & & OUT$^{55}$ \\		
J0402+826	&04 02 12.6736	&82 41 35.103& & & & & & &CSO (S)\\	     	     	     		
3C108	&04 12 43.6683	&23 05 05.468	&1293	&1000	&0.21	&0.6$_{v,f}$ & 1.215&	7& OUT \\
J0420+149	&04 20 51.0857	&14 59 15.634	&460	&310	&0.32&0.5$_3$ &  & & OUT$^{55}$  \\		
4C+68.05	&04 26 50.0654	&68 25 52.955&$<$454&	244&$<$0.50&0.6 &  & & ? \\		
J0431+206	&04 31 03.7585	&20 37 34.189	&3611	&2811	&0.20	&0.6$_s^1$ & 0.219&	41& CSO (S) \\
J0431+175	&04 31 57.3798	&17 31 35.792&	429	&270	&0.37& 0.4$_3$ & & & OUT$^{55}$ \\		
J0458+201	&04 58 29.8726	&20 11 35.997&	170	&163	&0.03&0.6$_{3,v}$ &  & & OBS\\		
4C+10.16	&05 16 46.6463	&10 57 54.773&	1207	&734	&0.40&0.5 &  1.580&143 & ?\\		
J0532+013	&05 32 08.7760	&01 20 06.330&	258	&153	&0.42&0.4$_3$ &  & & OUT$^{55}$ \\		
J0600+630	&06 00 27.0161	&63 04 07.481&$<$110&	114&$<-$0.02& & & &? \\		
J0626+621	&06 26 42.2118	&62 11 23.514&	195&	134	&0.30&0.6$_3$ &  & & ?\\		
J0639+351	&06 39 09.5887	&35 06 22.543&	346&	233&	0.32& 0.4 & & & ?\\		
J0641+356	&06 41 35.8542	&35 39 57.623&	340	&197&	0.44&0.5$^{0.1}$ &  & & MERLIN$^{55}$ \\		
J0653+646	&06 53 53.7227	&64 38 13.176&	178&	130&	0.25& 0.4$_3$ & & & ?\\		
J0654+427	&06 54 43.5263	&42 47 58.728&	188&	188&	0.00&0.2 & 	0.126&	25& OUT \\
J0656+321	&06 56 40.8892	&32 09 32.554&$<$110&	217&$<-$0.54& & & & OUT\\		
J0735+236	&07 35 59.9293	&23 41 02.764&	878&	552&	0.37&0.5 &  & & MERLIN/VLBI$^{55}$ \\		
J0751+826	&07 50 57.7640	&82 41 58.032&	1815& & & 	     	  0.5$_f$ &    	1.991&	10& MSO (t)\\
J0752+581	&07 52 09.6792	&58 08 52.256&	203&	212&	$-$0.03&0.0$_f$ & 	2.94&	3& OBS\\
J0757+611	&07 57 44.6933	&61 10 32.764&	246&	195&	0.19& 0.3$_{3,s}$ & & & ?\\		
J0803+640	&08 03 52.1595	&64 03 14.364&	292&	221&	0.22&0.2 &  & & ?\\		
J0815+019	&08 15 58.6371&	01 55 55.820&$<$110&	280&$<-$0.75&0.7 &  & & MERLIN \\ \hline		
\end{tabular}
\end{center}
\end{table*}
\begin{table*}
\begin{minipage}{\linewidth}
\begin{center}
\begin{tabular}{cccccccccc} \hline \hline
{\bf (1)} &  {\bf (2)} & {\bf (3)} & {\bf (4)} & {\bf (5)} &  {\bf (6)} &  {\bf (7)} &  {\bf (8)} &  {\bf (9)} &  {\bf (10)} \\
Name & RA (2000) & Dec (2000) & S$_{1.40}$ & S$_{4.85}$ & $\alpha_{1.40}^{4.85}$ & $\alpha_{thin}$ & z & Ref. & Notes \\
 &  &  & (mJy) & (mJy) & & & &  \\ \hline
J0817+324	&08 17 28.5455&	32 27 02.928&$<$564&	585&$<-$0.02&0.4$_{f}$ &  & & OUT \\		
J0817+556	&08 17 41.0199 &55 37 33.283  &	178	&244&	$-$0.25&0.3 &  & & ? \\		
J0822+708	&08 22 16.7649&	70 53 07.978&	436&	274&	0.37&0.7$^{0.1}$ &  & & MERLIN/VLBI$^{55}$ \\		
J0822+080	&08 22 33.1537&	08 04 53.523&$<$110&	204&$<-$0.49& & & &MERLIN/VLBI$^{55}$ \\		
J0824+392	&08 24 55.4837&	39 16 41.898&	1381&	1012&	0.25&0.5$_{f}$ & 	1.216&	9& OUT$^{55}$ \\
J0827+354	&08 27 38.5891&	35 25 05.081&	866	&746&	0.12&0.5$_{3,f}$ & 	2.249&	62& MERLIN$^{55}$ \\
J0832+278	&08 32 19.6581&	27 52 43.879&	467	&274&	0.43&0.5$_{3}$ &  & & OBS \\		
J0834+555	&08 34 54.9026&	55 34 21.086&	7741&	5780&	0.24&1.1$_{s}$ & 	0.242&	63&OUT$^{55}$  \\
J0837+584	&08 37 22.4100&	58 25 01.844&	597	&669&	$-$0.09&0$_{v}$ & 	2.101&	7& OUT \\
J0855+578	&08 55 21.3575&	57 51 44.082&$<$110&	279&$<-$0.74&0.7$_{3}$ &  & &VLBI \\		
J0901+671	&09 01 58.7485&	67 07 32.225&$<$219&	100&$<$0.64& & & &? \\		
J0908+418	&09 08 35.8623&	41 50 46.204&$<$110&	207&$<-$0.50&0.4$_{3}$ & 	0.7325&	73&OUT$^{55}$  \\
J0911+861	&09 11 37.7924	&86 07 33.504& & & &0.7$_{3}$ &  & & ?\\	     	     	     		
J0915+209	&09 15 08.7822&	20 56 07.367	&257&	191&	0.24&0.4$_{f}$ &  & & ?\\		
4C+58.18	&09 16 59.7887&	58 38 49.349&$<$1478&	313&$<$1.25&1.0 &  & & ?\\		
J0917+737	&09 17 28.0923&	73 43 13.460&	213&	161&	0.23&0.5$_{3}$ &  & & ? \\		
J0921+716	&09 21 23.9433&	71 36 12.417&	384	&292&	0.22&0.6$_{v}$ & 	0.594&	64&OUT$^{55}$  \\
3C225A	&09 42 08.4797&	13 51 54.229&$<$110&	376&$<-$0.98&0.0$_{3}$ & 	1.565&	81& OBS\\
4C+66.09	&09 49 12.1652&	66 14 59.587&	2223	&1407&	0.37&0.6$_{s}^{0.4}$ &  & & CSO/MSO (t) \\		
4C+55.17	&09 57 38.1825&	55 22 57.734&	3000&	2270	&0.22&0.4$_{v}$ & 	0.909&	7& OUT \\
J1002+122	&10 02 52.8457&	12 16 14.588&	179	&285&	$-$0.37&0.0$_{3,f}$ &  & & ?\\		
J1003+260	&10 03 42.2292&	26 05 12.903&	491&	274&	0.47&0.4$_{v}$ & 	& & ?\\	
J1005+240	&10 05 07.8712&	24 03 38.003&	209&	146&	0.29& & & &?\\		
J1006+172	&10 06 31.7650&	17 13 17.104&	497&	337&	0.31&0.5$_{3}$ &  & & OUT$^{55}$  \\		
J1013+284	&10 13 03.0002&	28 29 10.926&	612&	331&	0.49&0.6$_{3}^{0.1}$ &  & & MERLIN/VLBI$^{55}$ \\		
J1015+494	&10 15 04.1358&	49 26 00.692&	382&	286&	0.23&0.3 & 	0.2&	65& OUT$^{55}$ \\
J1015+674	&10 15 38.0161&	67 28 44.442&$<$110&	117 & $<-$0.05& & & & OUT\\		
J1034+594	&10 34 34.2393&	59 24 45.846&	164	&142&	0.12&0.5$_{3,f}$ &  (2.13069)\footnotemark &141 & OBS\\
J1035+568	&10 35 06.0207&	56 52 57.960&	273	&227&	0.15&0.6$_{3,f}$ & 	1.855420\footnotemark&	141& ?\\
\addtocounter{footnote}{-1}\footnotetext{${\rm \star \star}$ If the identification is a QSO $\sim9\arcsec$ away.}
\addtocounter{footnote}{1}\footnotetext{${\rm \dagger \dagger}$ The previous redshift measurement (z=0.577) by \protect \scite{Peretal98} is very different.}
\hspace*{-2mm}J1041+525	&10 41 46.7800&	52 33 28.217&	713&	709&	0.00&0.2 & 	0.677&	7& OUT \\
J1058+198	&10 58 17.8992&	19 51 50.902&	2310&	1678&	0.26&0.5$_{f}$ & 	1.11&	7& OUT\\
J1101+242	&11 01 23.5143&	24 14 29.517&	416&	231&	0.47	&0.6$_{3}^{0.1}$ &  & & MERLIN$^{55}$ \\	
J1108+020	&11 08 46.35&	02 02 43&	928&	678&	0.25&0.4$_{v}$ & 	0.157/0.158&	45,141,144& ?\\
J1110+440	&11 10 46.3458&	44 03 25.938&	373&	297&	0.18&0.2 &  & & OUT\\		
4C+20.25	&11 25 58.7440&	20 05 54.381&$<$110&	759&$<-$1.55&0.5$_{v}$ & 	0.133&	25& ?\\
J1132+005	&11 32 45.6189&	00 34 27.821&	472&	358&	0.22&0.5$_{v}$ &  1.22270&141 & ?\\		
J1141+497	&11 41 54.8254&	49 45 06.564&	157&	98&	0.38&0.6$_{3}$ &  & & ?\\		
J1145+443	&11 45 38.5190&	44 20 21.918&	438&	245&	0.47&0.3 & 	0.3&	22& OUT$^{55}$ \\
J1153+092	&11 53 12.5524&	09 14 02.312&	737&	500&	0.31&0.5$_{v}$ & 	0.698&	9& OUT$^{55}$ \\
J1159+583	&11 59 48.7733&	58 20 20.306&$<$1557&	369&$<$1.16	&0.7$_{3,s}$ &  & & ?\\	
J1213+131	&12 13 32.1412&	13 07 20.373&	1486&	894&	0.41&0.4&	1.141&	20& OBS\\
J1214+331	&12 14 04.1129&	33 09 45.556&	1196	&649&	0.49&0.3$_{s}^{0.4}$ & 	1.598&	9& OUT$^{55}$ \\
J1215+175	&12 15 14.7215&	17 30 02.250&	836&	620&	0.24&0.6$_{v}$ &  & & MERLIN/VLBI$^{55}$ \\		
J1224+435	&12 24 51.5059&	43 35 19.282&	393&	235&	0.41&0.3 &  1.872&145 & ?\\	
J1226+096	&12 26 25.4693&	09 40 04.432&	895&	524&	0.43&0.7& & & ?\\
J1235+536	&12 35 48.2529&	53 40 04.839&	392&	215&	0.48&0.5&1.97193 &141 & MERLIN$^{55}$ \\		
J1239+075	&12 39 24.5908&	07 30 17.225&	435&	674&	$-$0.35&0.0&	0.4&	7&OBS \\
J1243+732	&12 43 11.2156&	73 15 59.259&	518&	345&	0.33&0.6$_{s}$ & 	0.075&	25& OUT$^{55}$ \\
J1244+879	&12 44 06.7918&	87 55 08.093& & & &0.3$_{3}$ &  & & ?\\	     	     	     		
J1306+801	&13 06 05.7164&	80 08 20.543&	862& &	& &   1.183&	71& OUT\\
J1319+196	&13 19 52.0736&	19 41 35.481&	672&	3866	& 0.45&0.5$_{3}$ &  & & OUT$^{55}$ \\		
J1324+477	&13 24 29.3413&	47 43 20.624&	188&	237&	$-$0.19&$-0.1$ & 	2.26&	22&OUT \\
J1334+092	&13 34 19.5624&	09 12 00.366&	360&	266&	0.24&0.4$_{3}$ &  & & ?\\		
J1344+339	&13 44 37.1019&	33 55 46.195&	262&	154&	0.43&0.4& & & OUT$^{55}$ \\		
J1344+791	&13 44 55.7307&	79 07 10.834&	370& & & & & &?\\	     	     		
J1411+592	&14 11 21.9856&	59 17 04.302&	326&	184&	0.46&0.6&	1.725&	73& ?\\
J1424+229	&14 24 38.0940&	22 56 00.590&	220&	503&	$-$0.67&	&3.62&     103& OUT \\ 
J1437+636	&14 37 41.3537&	63 40 05.772&$<$110&	237&$<-$0.61& & & &?\\						
J1440+383	&14 40 22.3365&	38 20 13.627&	1025&	944&	0.07&0.1&	1.775&	71& OUT\\
	    \hline  		
\end{tabular}
\end{center}
\end{minipage}
\end{table*}
\begin{table*}
\begin{minipage}{\linewidth}
\def\footnoterule{\kern-3pt
\hrule width 2truein height 0pt\kern3pt}
\begin{center}
\begin{tabular}{cccccccccc} \hline \hline
{\bf (1)} &  {\bf (2)} & {\bf (3)} & {\bf (4)} & {\bf (5)} &  {\bf (6)} &  {\bf (7)} &  {\bf (8)} &  {\bf (9)} &  {\bf (10)} \\
Name & RA (2000) & Dec (2000) & S$_{1.40}$ & S$_{4.85}$ & $\alpha_{1.40}^{4.85}$ & $\alpha_{thin}$ & z & Ref. & Notes \\
 &  &  & (mJy) & (mJy) & & & &  \\ \hline
4C+13.53	&14 42 04.0423&	13 29 16.067&	545&	341&	0.38&0.6& & & ?\\ 					
J1454+299	&14 54 32.3006&	29 55 58.110&	822&	460&	0.47&0.5$_{f}$ & 	0.58&	7&MSO (t) \\
J1504+689	&15 04 12.7748&	68 56 12.830& &	     	229& & & 0.318& 7& OUT\\
J1507+103	&15 07 21.8815&	10 18 44.988&	368&	227&	0.39&0.5$_{3}$ &  & & MERLIN/VLBI$^{55}$ \\		
J1526+099	&15 26 46.3484&	09 59 10.538&	430&	346&	0.17&0.7&	1.358&	7& OUT\\
J1530+059	&15 30 28.4444&	05 55 13.030&	552&	308&	0.47&0.6$_{3}$ &  & & OBS\\		
Arp220	&15 34 57.2240&	23 30 11.608&	302&	204&	0.32&0.3$_{s}^{0.1}$ & 	0.018126& 86& OUT\\
4C+49.26	&15 47 21.1384&	49 37 05.810&	936&	549&	0.43&0.5$_{s}$ & 	0.7&	22& OUT \\
J1607+158	&16 07 06.4309&	15 51 34.503&	603&	512&	0.13&0.6$_{f}$ & 	0.357&	113& OUT \\
J1617+041	&16 17 13.5894&	04 08 41.674&	372&	439&	$-$0.13&0.3$_{3}$ &  & & ?\\		
J1630+215	&16 30 11.2359&	21 31 34.379&	300&	245&	0.16&0.6$_{3,v}$ & 	& & MERLIN$^{55}$ \\	
4C+12.59	&16 31 45.2469&	11 56 02.991&	1628&	954&	0.43&0.6&	1.795&	7& OUT\\
J1635+599	&16 35 37.6511&	59 55 15.097&	234&	218&	0.06& & & & ?\\		
J1640+123	&16 40 47.9384&	12 20 02.108&	2066&	1292&	0.38&0.6$^{0.3}$ & 	1.152&	66& OUT$^{55}$ \\
J1644+053	&16 44 56.0829&	05 18 37.064&	659&	393&	0.42&0.5$_{3}$ &  & & OUT$^{55}$ \\		
J1706+523	&17 06 00.9421&	52 18 42.748&	     &	164& & & & &? \\	     		
J1713+492	&17 13 35.1484&	49 16 32.548&	260&	216&	0.15&0.4$_{f}$ & 	1.552&	73& ?\\
J1715+217	&17 15 21.2517&	21 45 31.709&	580&	327&	0.46&0.5$_{3}$ & 	4.011&	104& VLBI\\
J1722+561	&17 22 58.0083&	56 11 22.320&	219&	132&	0.41&0.7& & & OUT$^{55}$ \\		
J1746+260	&17 46 48.2909&	26 03 20.343&	385&	261&	0.31&0.3$_{v}$ & 	0.147&	25&OUT$^{55}$  \\
J1749+431	&17 49 00.3604&	43 21 51.289&	340&	367&	$-$0.06&0.2& & &  OUT\\		
J1751+509	&17 51 32.5892&	50 55 37.847&	310&	192&	0.39&0.6&	0.3284&   73&OBS \\
J1752+455	&17 52 26.1411&	45 30 59.120&$<$223&	104&$<$0.62& & & & ?\\		
J1753+093	&17 53 02.5264&	09 20 02.982&	883&	514&	0.44&0.5$_{f}$ &  0\footnotemark & & OUT \\		
\addtocounter{footnote}{0}\footnotetext{${\rm \ddagger \ddagger}$ This is a galactic star (see text).}		
NGC6521	&17 55 48.4397&	62 36 44.119&$<$110&	198&$<-$0.47&0.6$_{3}$ & 	0.027462& 108&OUT \\
J1755+049	&17 55 51.1535&	04 54 52.566&	     	&244& & & & &?\\	     		
J1759+464	&17 59 41.7970&	46 27 59.906&$<$110&	124&$<-$0.09& & & &? \\		
J1803+036	&18 03 56.2829&	03 41 07.575&	     	&250& & & & & CSO? (S)\\	     		
NGC6572	&18 12 06.2100&	06 51 13.382&	495&	1251	&$-$0.75& & 0\footnotemark & & OUT \\		
\addtocounter{footnote}{0}\footnotetext{${\rm \S \S}$ This is a galactic planetary nebulae (see text).}
\hspace*{-1mm}J1814+412	&18 14 22.7082&	41 13 05.605&	644	&534&	0.15&0.4&	1.564&	23&OUT$^{55}$  \\
J1829+399	&18 29 56.5203&	39 57 34.690&	127&	353&	$-$0.82& & & &?\\		
J1857+630	&18 57 29.1989&	63 05 30.043&	263&	164&	0.38&0.6& & &OUT$^{55}$  \\		
J1928+682	&19 28 20.5502&	68 14 59.247&	533&	319&	0.41&0.5$^{0.3}$ &  & & CSO? (S) \\		
J1947+678	&19 47 36.2599&	67 50 16.928&	264&	165&	0.38&0.5$_{3}$ &  & & CSO? (S) \\		
J2007+748	&20 07 04.3881&	74 52 25.398&	283&	262&	0.06&0.2$_{3}$ &  & & ?\\		
J2035+583	&20 35 23.7535&	58 21 18.759&	313&	220&	0.28&0.3$_{s}$ &  & &OBS \\		
J2045+741	&20 45 42.8810&	74 09 54.800&	223&	128&	0.45& & & &?\\		
J2055+287	&20 55 30.5466&	28 47 38.347& &	     	232& &0.7$_{3}$ &  & & MSO? (t)\\	     		
J2102+666	&21 02 36.6376&	66 36 34.217&	&     	76& & & & &OUT$^{55}$ \\	     		
J2114+315	&21 14 50.4610&	31 30 21.183&	440&	255&	0.44&0.5$_{3}$ &  & & OUT$^{55}$ \\		
J2144+190	&21 44 57.7115&	19 05 18.945&	305&	222&	0.26&0.5$_{3}$ &  & & ?\\		
J2153+126	&21 53 04.6587&	12 41 05.211&	422&	264&	0.38&0.5$_{3}$ &  2.22278&141 & OUT$^{55}$ \\		
J2153+176	&21 53 36.8267&	17 41 43.726&	282&	241&	0.13&0.3&	0.231&	67& MSO (S)\\
J2204+046	&22 04 17.6522	&04 40 02.007	&784&	747&	0.04&0.5$_{v,f}$ & 	0.028&	9& OUT$^{55}$ \\
J2207+392	&22 07 46.0720&	39 13 50.353&	445&	294&	0.33&0.6& & & OUT$^{55}$ \\		
J2213+087	&22 13 21.7374&	08 47 29.951&	226&	208&	0.07&0.4$_{3,f}$ &  & & OUT$^{55}$ \\		
J2217+204	&22 17 15.8391&	20 24 48.970&	408&	221&	0.49&0.6$_{3}$ &  & &? \\		
J2234+361	&22 34 02.9764&	36 11 00.333& &	     	141 & & & & &OUT\\ 
J2250+143	&22 50 25.3434&	14 19 52.044&	2127&	1177&	0.48&0.4 &	0.237	&68&OUT$^{55}$  \\
J2344+278	&23 44 37.0573&	27 48 35.521&$<$110&	148&$<-$0.23&	&0.0573&	37& ?\\
J2347+115	&23 47 36.4062&	11 35 17.893&	313&	201&	0.36&0.5$_{3}$ &  & & MERLIN$^{55}$ \\ \hline
\end{tabular}
\end{center}
\end{minipage}
\end{table*}

\subsection{Statistics}

\begin{figure} 
\setlength{\unitlength}{1cm}
\begin{picture}(8,6.5)
  \put(-.5,7.2){\includegraphics{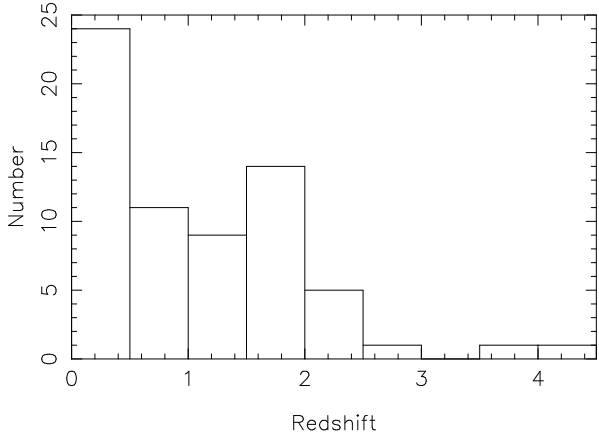}}
\end{picture}
\caption{ The redshift distribution of the 66 sources of the 157-source sample (minus two galactic sources) which have such information (43\% completeness).} 
\label{zdist_157}
\end{figure}

\begin{figure} 
\setlength{\unitlength}{1cm}
\begin{picture}(8,6.5)
  \put(-0.5,7.2){\includegraphics{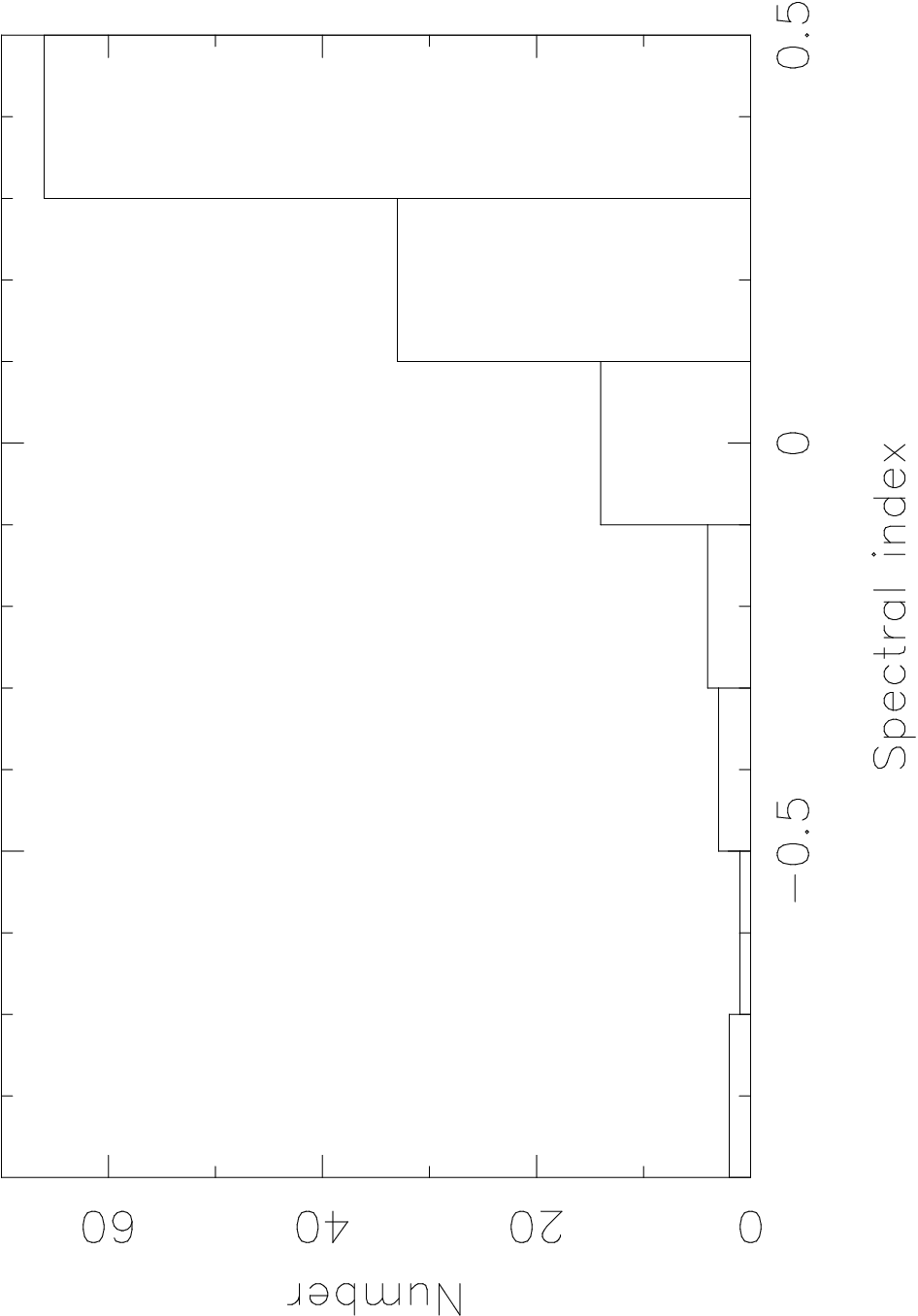}}
\end{picture}
\caption{ The 1.40--4.85~GHz spectral index distribution of the 123 sources of the 157-source parent sample which have such data (78\% completeness).} 
\label{a_dist_157}
\end{figure}

In Figures~\ref{zdist_157} and~\ref{a_dist_157} we plot, respectively, the redshift ($z$) and spectral index ($\alpha_{1.40}^{4.85}$) distributions for the 157-source sample (which have different completenesses). The compared statistics of this new sample with the previous 55-source sample of \scite{papI} are discussed in Appendix~A. Relevant here is the comparison  with the new 1743-source parent sample
(Figure~\ref{zdist} vs. \ref{zdist_157}; Figure~\ref{a_dist} vs. \ref{a_dist_157}),
with results shown in Table~6.
The completenesses are similar for both samples. This similarity, in the redshift case, is not surprising since the two samples have the same  flux density lower limit. The completeness similarity in the case of the spectral index merely reflects that we are not biasing our selection towards ``better known'' or brighter sources (which is good, since we want a morphological-only difference): the proportion of sources that are too weak to be found on one (or both) of the \scite{WhiBec92} and \scite{GreCon91}  catalogues is the same.

\begin{table*}
\begin{center}
\label{compza}
\caption{Comparison of the redshift ($z$) and spectral index distributions ($\alpha_{1.40}^{4.85}$) between the parent sample and the 157-source sample. The medians are indicated with their asymmetric errors (95\% conf. level). We also give the number of sources for this calculation in each case (\#) and the correspondent completeness to the whole sample. Finally, we indicate the Figures where the histograms are plotted.}
\begin{tabular}{ccccccccc}
{\it sample} & $z$ (median) & \# & comp.\ & Fig.\ & $\alpha_{1.40}^{4.85}$ & \# & comp.\ & Fig.\ \\ \hline
parent & $1.12^{+0.11}_{-0.10}$ & 675 & 39\% & 11 & $0.08^{+0.03}_{-0.01}$ & 1311 & 75\% & 12 \\
157-source & $0.88^{+0.34}_{-0.48}$ & 66 & 43\% & 13 & $0.32^{+0.05}_{-0.07}$ & 123 & 78\% & 14  \\\hline
\end{tabular}
\end{center}
\end{table*}

As regards as the redshift distributions,  the difference is obvious by eye, with the 157-source sample containing more low-z sources than the parent sample. Furthermore, the latter has a smooth redshift distribution, roughly flattening around $z\sim0.7$ and having an average redshift coincident with that of other flat spectrum radio source samples ($<\!\!z\!\!>\;\simeq1.2$; \pcite{Munetal03}). The median values also suggest a selection of the closest radio sources (0.88 vs. 1.12), albeit with intersecting values, within the 95\% conf. level errors.
 More formally, we have applied a KS-test to compare the two distributions and reject the hypothesis that they are similar at the 95\% confidence level. 

As for the spectral indices distributions we have, again, a smooth distribution for the parent sample (roughly flattening at $\alpha_{1.40}^{4.85}\simeq0.2$) while for the 157-source sample the distribution is also smooth but still rising when it reaches the limit of $\alpha_{1.40}^{4.85}=0.5$. This time, the distributions are clearly different by eye and medians (whose errors do not overlap). We performed the formal KS-test to compare both distributions and rejected the hypothesis that they are similar at the 99.9\% confidence level. This result is similar to the one of \scite{papI} and also similarly explained by the fact that we are selecting resolved sources from the parent sample and this (normally) implies steeper spectrum sources.

\section{Summary}

In what follows we briefly summarize the main conclusions from this paper:

\begin{itemize}
\item In order to understand the origin and evolution of   extragalactic radio sources in the context of the standard model of AGN, several VLBI searches have been conducted trying to identify the youngest such sources ($\la10^3$~yrs), of which compact symmetric objects (CSOs) are the most serious contenders. Ideally, we should also follow the evolutionary track at later stages, by identifying somewhat older sources ($\sim10^4$--10$^5$~yrs), possibly medium symmetric objects (MSOs). In this paper we summarize all confirmed cases of CSOs that we found from the literature, which total to 37, and three $\alpha_{1.40}^{4.85}<0.5$ flat spectrum MSOs. 
\item By studying the sample of the currently confirmed 37 CSOs we conclude the following (the completeness of the statistics is $\geq62\%$ but beware that the sample might not be representative of the CSO class due to the heterogeneous surveys from where the sources were selected): {\bf i)} 85\% of the optical hosts are galaxies, typically residing at $z\la0.5$; the remaining are quasars, with a large spread in redshift range; {\bf ii)} most CSOs have flat radio spectra (70\% with $\alpha_{1.40}^{4.85}<0.5$; 56\% with $\alpha_{thin}<0.5$); {\bf iii)} most (17, 59\%) CSOs follow the ``classical'' \cite{Wiletal94,Conetal94} definition where the brightness of the nucleus is $<10\%$ of the one of the brightest lobe; one-third of the CSOs present nuclear components that are brighter than the brightest of the two opposed lobes --- is this evidence for boosting?; 
 {\bf iv)} all CSO/MSOs were defined to have arm length ratios $R\leq10$ (for symmetry); the maximum value on the present sample is $R=4.6$, with 90\% (all but three) having $R\leq3.0$; {\bf v)} 73\% of the CSO/MSOs also present symmetry in the flux density ratios between the two lobes ($\leq10$); however, this ratio can be as large as 113, among the remaining;
{\bf vi)} 76\% of CSO/MSOs have well aligned opposing structures ($\theta\leq20^{\circ}$) but values as large as $\theta=46^{\circ}$ can be found; {\bf vii)} CSOs have a median linear projected size of $0.14^{+0.07}_{-0.05}$~kpc, with 84\% smaller than 0.3~kpc.
\item The aim of the series of papers starting with this one is to improve, many times, the number of confirmed large CSOs and of flat-spectrum MSOs (0.3--15~kpc), which currently sits at six. In this paper, in particular, we present a sample of 157 sources, drawn from a parent sample of 1743 flat spectrum ($\alpha_{1.40}^{4.85}<0.5$) sources by selecting the ones with radio structure on $\ga0\farcs1$ scales. This resulted in the selection of the lowest redshift and steepest spectrum sources including $\ga0.3$~kpc CSO/MSO candidates.
 Although we have immediately rejected, based on literature information, 61 of the sources, 83 are still left with data either to be analysed or to be gathered. As for the remaining thirteen sources, nine were already listed as CSOs/flat-spectrum MSOs from the literature and are, thus, a good quality control for our selection. As for the final four, 4C+66.09 is a CSO/MSO (needs  a redshift to identify which type exactly); J0751+826 and J1454+299  are $\sim14$~kpc MSOs; J2055+287  might be  an MSO too, if at $z<0.09$.
\end{itemize}

\section*{Acknowledgments}

The authors acknowledge an anonymous referee, whose comments helped to improve this paper, and also support from the Funda\c{c}\~ao para a
Ci\^encia e a Tecnologia (FCT) under the ESO programme: PESO/P/PRO/15133/1999. We thank Daniele Dallacasa for information on one source, before publishing. 
This paper has made use of the NASA/IPAC Extragalactic Database (NED) which is operated by the Jet Propulsion Laboratory, California Institute of Technology, under contract with the National Aeronautics and Space Administration. 
Funding for the Sloan Digital Sky Survey (SDSS) has been provided by the Alfred P. Sloan Foundation, the Participating Institutions, the National Aeronautics and Space Administration, the National Science Foundation, the U.S. Department of Energy, the Japanese Monbukagakusho, and the Max Planck Society. The SDSS Web site is http://www.sdss.org/.
The SDSS is managed by the Astrophysical Research Consortium (ARC) for the Participating Institutions. The Participating Institutions are The University of Chicago, Fermilab, the Institute for Advanced Study, the Japan Participation Group, The Johns Hopkins University, the Korean Scientist Group, Los Alamos National Laboratory, the Max-Planck-Institute for Astronomy (MPIA), the Max-Planck-Institute for Astrophysics (MPA), New Mexico State University, University of Pittsburgh, University of Portsmouth, Princeton University, the United States Naval Observatory, and the University of Washington.


 \nocite{HerRea92,FeyCleFom96,Keletal98,Galetal99,Staetal97b,Cotetal95,Zenetal02,Fometal00,Tayetal94,Daletal02a,Daletal02b,Caretal98,Gioetal01,SneSchvLa00a,Staetal97,StiKuh93a,StaODeMur99,AkuPorSmo96,Conetal92,AllAllHug92,deVBarHes95,Ungetal84,WikCom97,Peretal94,Peretal96,Tadetal93,Tzietal02,Ojhetal04,Baretal84,Syk97,Orietal04,Gugetal05,Pecetal00,Miletal02,GirGioTay05,OwsConPol98,TayPec03,Tayetal00,Wil95}

\bibliography{Papers}

\begin{thebibliography}{Kunert-Bajraszewska {\rm et~al.}<2005>}
\bibitem[Akujor {\rm et~al.}<1996>]{AkuPorSmo96}
Akujor~C.~E., Porcas~R.~W., Smoker~J.~V., 1996, A\&A, 306, 391
\bibitem[Alexander<2000>]{Ale00}
Alexander~P., 2000, MNRAS, 319, 8
\bibitem[Aller {\rm et~al.}<1992>]{AllAllHug92}
Aller~M.~F., Aller~H.~D., Hughes~P.~A., 1992, ApJ, 399, 16
\bibitem[Allington-Smith {\rm et~al.}<1988>]{Alletal88}
Allington-Smith~J.~R., Spinrad~H., Djorgovski~S., Liebert~J., 1988, MNRAS, 234,
  1091
\bibitem[Anton {\rm et~al.}<2002>]{Antetal02}
Anton~S., Thean~A. H.~C., Pedlar~A., Browne~I. W.~A., 2002, MNRAS, 336, 319
\bibitem[Augusto \& Wilkinson<2001>]{AugWil01}
Augusto~P., Wilkinson~P.~N., 2001, MNRAS, 320, L40
\bibitem[Augusto {\rm et~al.}<1998>]{papI}
Augusto~P., Wilkinson~P.~N., Browne~I. W.~A., 1998, MNRAS, 299, 1159
\bibitem[Augusto {\rm et~al.}<2005>]{Augetal04}
Augusto~P., Edge~A.~C., Chandler~C.~J., 2005, MNRAS, submitted
\bibitem[Augusto<1996>]{Aug96}
Augusto~P., 1996, Ph.D. Thesis, University of Manchester, UK
\bibitem[Baldwin {\rm et~al.}<1973>]{Baletal73}
Baldwin~J.~A., Burbidge~E.~M., Hazard~C., Murdoch~H.~S., Robinson~L.~B.,
  Wampler~E.~J., 1973, ApJ, 185, 739
\bibitem[Bartel {\rm et~al.}<1984>]{Baretal84}
Bartel~N. {\rm et~al.}, 1984, ApJ, 279, 116
\bibitem[Baum {\rm et~al.}<1990>]{Bauetal90}
Baum~S.~A., O'Dea~C.~P., de~Bruyn~A.~G., Murphy~D.~W., 1990, A\&A, 232, 19
\bibitem[Beasley {\rm et~al.}<2002>]{Beaetal02}
Beasley~A.~J., Gordon~D., Peck~A.~B., Petrov~L., MacMillan~D.~S.,
  Fomalont~E.~B., Ma~C., 2002, ApJS, 141, 13
\bibitem[Begelman<1996>]{Beg96}
Begelman~M.~C., 1996, in Carilli~C.~L., Harris~D.~E., eds, Cygnus A --- Study
  of a Radio Galaxy (Proc. of Greenbank workshop).
\newblock Cambridge University Press, p.~209
\bibitem[Blandford \& Rees<1974>]{BlaRee74}
Blandford~R.~D., Rees~M.~J., 1974, MNRAS, 169, 395
\bibitem[Bondi {\rm et~al.}<1998>]{BonGarGur98}
Bondi~M., Garrett~M.~A., Gurvits~L.~I., 1998, MNRAS, 297, 559
\bibitem[Bondi {\rm et~al.}<2001>]{Bonetal01}
Bondi~M., Marcha~M. J.~M., Dallacasa~D., Stanghellini~C., 2001, MNRAS, 325,
  1109
\bibitem[Burbidge \& Crowne<1979>]{BurCro79}
Burbidge~G., Crowne~A.~H., 1979, ApJS, 40, 583
\bibitem[Carilli {\rm et~al.}<1998>]{Caretal98}
Carilli~C.~L., Menten~K.~M., Reid~M.~J., Rupen~M.~P., Yun~M.~S., 1998, ApJ,
  494, 175
\bibitem[Carvalho<1985>]{Car85}
Carvalho~J.~C., 1985, MNRAS, 215, 463
\bibitem[Clements {\rm et~al.}<2002>]{Cleetal02}
Clements~D.~L., McDowell~J.~C., Shaked~S., Baker~A.~C., Borne~K., Colina~L.,
  Lamb~S.~A., Mundell~C., 2002, ApJ, 581, 974
\bibitem[Cohen {\rm et~al.}<2003>]{CohLawBla03}
Cohen~J.~G., Lawrence~C.~R., Blandford~R.~D., 2003, ApJ, 583, 67
\bibitem[Condon \& Kaplan<1998>]{ConKap98}
Condon~J.~J., Kaplan~D.~L., 1998, ApJS, 117, 361
\bibitem[Condon {\rm et~al.}<1998>]{Conetal98}
Condon~J.~J., Cotton~W.~D., Greisen~E.~W., Yin~Q.~F., Perley~R.~A.,
  Taylor~G.~B., Broderick~J.~J., 1998, AJ, 115, 1693
\bibitem[Condon {\rm et~al.}<2002>]{ConCotBro02}
Condon~J.~J., Cotton~W.~D., Broderick~J.~J., 2002, AJ, 124, 675
\bibitem[Conway {\rm et~al.}<1992>]{Conetal92}
Conway~J.~E., Pearson~T.~J., Readhead~A. C.~S., Unwin~S.~C., Xu~W.,
  Mutel~R.~L., 1992, ApJ, 396, 62
\bibitem[Conway {\rm et~al.}<1994>]{Conetal94}
Conway~J.~E., Myers~S.~T., Pearson~T.~J., Readhead~A. C.~S., Unwin~S.~C.,
  Xu~W., 1994, ApJ, 425, 568
\bibitem[Cotton {\rm et~al.}<1995>]{Cotetal95}
Cotton~W.~D., Feretti~L., Giovannini~G., Venturi~T., Lara~L., Marcaide~J.,
  Wehrle~A.~E., 1995, ApJ, 452, 605
\bibitem[Dallacasa {\rm et~al.}<1995>]{Daletal95}
Dallacasa~D., Fanti~C., Fanti~R., Schilizzi~R.~T., Spencer~R.~E., 1995, A\&A,
  295, 27
\bibitem[Dallacasa {\rm et~al.}<1998>]{Daletal98}
Dallacasa~D., Bondi~M., Alef~W., Mantovani~F., 1998, A\&AS, 129, 219
\bibitem[Dallacasa {\rm et~al.}<2002a>]{Daletal02a}
Dallacasa~D., Tinti~S., Fanti~C., Fanti~R., Gregorini~L., Stanghellini~C.,
  Vigotti~M., 2002a, A\&A, 389, 115
\bibitem[Dallacasa {\rm et~al.}<2002b>]{Daletal02b}
Dallacasa~D., Fanti~C., Giacintucci~S., Stanghellini~C., Fanti~R.,
  Gregorini~L., Vigotti~M., 2002b, A\&A, 389, 126
\bibitem[de~Vaucouleurs {\rm et~al.}<1991>]{deVetal91}
de~Vaucouleurs~G., de~Vaucouleurs~A., Corwin~Jr.~H.~G., Buta~R.~J., Paturel~G.,
  Fouqu\'{e}~P., 1991, in ``Third Reference Catalogue of Bright Galaxies''.
\newblock Springer-Verlag
\bibitem[de~Vries {\rm et~al.}<1995>]{deVBarHes95}
de~Vries~W.~H., Barthel~P.~D., Hes~R., 1995, A\&AS, 114, 259
\bibitem[de~Vries {\rm et~al.}<1997>]{deVetal97}
de~Vries~W.~H. {\rm et~al.}, 1997, ApJS, 110, 191
\bibitem[de~Vries {\rm et~al.}<1998a>]{deVetal98a}
de~Vries~W.~H., O'Dea~C.~P., Perlman~E., Baum~S.~A., Lehnert~M.~D., Stocke~J.,
  Rector~T., Elston~R., 1998a, ApJ, 503, 138
\bibitem[de~Vries {\rm et~al.}<1998b>]{deVetal98b}
de~Vries~W.~H., O'Dea~C.~P., Baum~S.~A., Perlman~E., Lehnert~M.~D.,
  Barthel~P.~D., 1998b, ApJ, 503, 156
\bibitem[de~Vries {\rm et~al.}<2000>]{deVetal00}
de~Vries~W.~H., O'Dea~C.~P., Barthel~P.~D., Fanti~C., Fanti~R., Lehnert~M.~D.,
  2000, AJ, 120, 2300
\bibitem[de~Young<1997>]{DeY97}
de~Young~D.~S., 1997, ApJ, 490, L55
\bibitem[Dondi \& Ghisellini<1995>]{DonGhi95}
Dondi~L., Ghisellini~G., 1995, MNRAS, 273, 583
\bibitem[Drinkwater {\rm et~al.}<1997>]{Drietal97}
Drinkwater~M.~J. {\rm et~al.}, 1997, MNRAS, 284, 85
\bibitem[Emerson {\rm et~al.}<1984>]{Emeetal84}
Emerson~J.~P., Clegg~P.~E., Gee~G., Griffin~M.~J., Cunningham~C.~T., Brown~L.
  M.~J., Robson~E.~I., Longmore~A.~J., 1984, Nature, 311, 237
\bibitem[Falco {\rm et~al.}<1998>]{FalKocMun98}
Falco~E.~E., Kochanek~C.~S., Munoz~J.~A., 1998, ApJ, 494, 47
\bibitem[Fanaroff \& Riley<1974>]{FanRil74}
Fanaroff~B.~L., Riley~J.~M., 1974, MNRAS, 167, P31
\bibitem[Fanti {\rm et~al.}<1990>]{Fanetal90}
Fanti~R., Fanti~C., Schilizzi~R.~T., Spencer~R.~E., Nan~R., Parma~P., van
  Breugel~W. J.~M., Venturi~T., 1990, A\&A, 231, 333
\bibitem[Fanti {\rm et~al.}<1995>]{Fanetal95}
Fanti~C., Fanti~R., Dallacasa~D., Schilizzi~R.~T., Spencer~R.~E.,
  Stanghellini~C., 1995, A\&A, 302, 317
\bibitem[Fanti {\rm et~al.}<2001>]{Fanetal01}
Fanti~C., Pozzi~F., Dallacasa~D., Fanti~R., Gregorini~L., Stanghellini~C.,
  Vigotti~M., 2001, A\&A, 369, 380
\bibitem[Fassnacht \& Taylor<2001>]{FasTay01}
Fassnacht~C.~D., Taylor~G.~B., 2001, AJ, 122, 1661
\bibitem[Fey {\rm et~al.}<1996>]{FeyCleFom96}
Fey~A.~L., Clegg~A.~W., Fomalont~E.~B., 1996, ApJS, 105, 299
\bibitem[Fomalont {\rm et~al.}<2000>]{Fometal00}
Fomalont~E.~B., Frey~S., Paragi~Z., Gurvits~L.~I., Scott~W.~K., Taylor~A.~R.,
  Edwards~P.~G., Hirabayashi~H., 2000, ApJS, 131, 95
\bibitem[Gallimore {\rm et~al.}<1999>]{Galetal99}
Gallimore~J.~F., Baum~S.~A., O'Dea~C.~P., Pedlar~A., Brinks~E., 1999, ApJ, 524,
  684
\bibitem[Giovannini {\rm et~al.}<2001>]{Gioetal01}
Giovannini~G., Cotton~W.~D., Feretti~L., Lara~L., Venturi~T., 2001, ApJ, 552,
  508
\bibitem[Giroletti {\rm et~al.}<2003>]{Giretal03}
Giroletti~M., Giovannini~G., Taylor~G.~B., Conway~J.~E., Lara~L., Venturi~T.,
  2003, A\&A, 399, 889
\bibitem[Giroletti {\rm et~al.}<2005>]{GirGioTay05}
Giroletti~M., Giovannini~G., Taylor~G.~B., 2005, A\&A, 441, 89
\bibitem[Goncalves {\rm et~al.}<1998>]{GonVerVer98}
Goncalves~A.~C., Veron~P., Veron-Cetty~M.-P., 1998, A\&AS, 127, 107
\bibitem[Gorshkov {\rm et~al.}<2003>]{GorKonMin03}
Gorshkov~A.~G., Konnikova~V.~K., Mingaliev~M.~G., 2003, Astron. Rep., 47, 903
\bibitem[Graham {\rm et~al.}<1990>]{Graetal90}
Graham~J.~R., Carico~D.~P., Matthews~K., Neugebauer~G., Soifer~B.~T.,
  Wilson~T.~D., 1990, ApJ, 354, L5
\bibitem[Gregory \& Condon<1991>]{GreCon91}
Gregory~P.~C., Condon~J.~J., 1991, ApJS, 75, 1011
\bibitem[Gregory {\rm et~al.}<1996>]{Greetal96}
Gregory~P.~C., Scott~W.~K., Douglas~K., Condon~J.~J., 1996, ApJS, 103, 427
\bibitem[Gugliucci {\rm et~al.}<2005>]{Gugetal05}
Gugliucci~N.~E., Taylor~G.~B., Peck~A.~B., Giroletti~M., 2005, ApJ, 622, 136
\bibitem[Heckman {\rm et~al.}<1996>]{Hecetal96}
Heckman~T.~M., Dahlem~M., Eales~S.~A., Fabbiano~G., Weaver~K., 1996, ApJ, 457,
  616
\bibitem[Henstock {\rm et~al.}<1995>]{Henetal95}
Henstock~D.~R., Browne~I. W.~A., Wilkinson~P.~N., Taylor~G.~B.,
  Vermeulen~R.~C., Pearson~T.~J., Readhead~A. C.~S., 1995, ApJS, 100, 1
\bibitem[Henstock {\rm et~al.}<1997>]{Henetal97}
Henstock~D.~R., Browne~I. W.~A., Wilkinson~P.~N., McMahon~R.~G., 1997, MNRAS,
  290, 380
\bibitem[Herbig \& Readhead<1992>]{HerRea92}
Herbig~T., Readhead~A. C.~S., 1992, ApJS, 81, 83
\bibitem[Hewett {\rm et~al.}<1995>]{HewFolCha95}
Hewett~P.~C., Foltz~C.~B., Chaffee~F.~H., 1995, AJ, 109, 1498
\bibitem[Hewitt \& Burbidge<1989>]{HewBur89}
Hewitt~A., Burbidge~G., 1989, ApJS, 63, 1
\bibitem[Hewitt \& Burbidge<1991>]{HewBur91}
Hewitt~A., Burbidge~G., 1991, ApJS, 75, 297
\bibitem[Hintzen {\rm et~al.}<1983>]{HinUlvOwe83}
Hintzen~P., Ulvestad~J., Owen~F., 1983, AJ, 88, 709
\bibitem[Hooimeyer {\rm et~al.}<1992>]{Hooetal92}
Hooimeyer~J. R.~A., Schilizzi~R.~T., Miley~G.~K., Barthel~P.~D., 1992, A\&A,
  261, 25
\bibitem[Hook \& McMahon<1998>]{HooMcM98}
Hook~I.~M., McMahon~R.~G., 1998, MNRAS, 294, L7
\bibitem[Hook {\rm et~al.}<1996>]{Hooetal96}
Hook~I.~M., McMahon~R.~G., Irwin~M.~J., Hazard~C., 1996, MNRAS, 282, 1274
\bibitem[Kaiser \& Alexander<1997>]{KaiAle97}
Kaiser~C.~R., Alexander~P., 1997, MNRAS, 286, 215
\bibitem[Kellerman {\rm et~al.}<1998>]{Keletal98}
Kellerman~K.~I., Vermeulen~R.~C., Zensus~J.~A., Cohen~M.~H., 1998, AJ, 115,
  1295
\bibitem[Kunert-Bajraszewska {\rm et~al.}<2005>]{Kunetal05}
Kunert-Bajraszewska~M., Marecki~A., Thomasson~P., Spencer~R.~E., 2005, A\&A,
  440, 93
\bibitem[Kunert {\rm et~al.}<2002>]{Kunetal02}
Kunert~M., Marecki~A., Spencer~R.~E., Kus~A.~J., Niezgoda~J., 2002, A\&A, 391,
  47
\bibitem[Lara {\rm et~al.}<2001>]{Laretal01}
Lara~L., Cotton~W.~D., Feretti~L., Giovannini~G., Marcaide~J.~M., Marquez~I.,
  Venturi~T., 2001, A\&A, 370, 409
\bibitem[Le~Borgne {\rm et~al.}<1991>]{LeBetal91}
Le~Borgne~J.-F., Mathez~G., Mellier~Y., Pell\'{o}~R., Sanahuja~B., Soucail~G.,
  1991, A\&AS, 88, 133
\bibitem[Lonsdale {\rm et~al.}<1998>]{LonBarMil93}
Lonsdale~C.~J., Barthel~P.~D., Miley~G.~K., 1998, ApJS, 87, 63
\bibitem[Magliocchetti {\rm et~al.}<2004>]{Magetal04}
Magliocchetti~M. {\rm et~al.}, 2004, MNRAS, 350, 1485
\bibitem[Marcha {\rm et~al.}<1996>]{Maretal96}
Marcha~M. J.~M., Browne~I. W.~A., Impey~C.~D., Smith~P.~S., 1996, MNRAS, 281,
  425
\bibitem[Marecki {\rm et~al.}<1999>]{Maretal99}
Marecki~A., Falcke~H., Niezgoda~J., Garrington~S.~T., Patnaik~A.~R., 1999,
  A\&AS, 135, 273
\bibitem[Marecki {\rm et~al.}<2003>]{MarSpeKun03}
Marecki~A., Spencer~R.~E., Kunert~M., 2003, PASA, 20, 46
\bibitem[Middelberg {\rm et~al.}<2004>]{Midetal04}
Middelberg~E. {\rm et~al.}, 2004, A\&A, 417, 925
\bibitem[Miller \& Owen<2001>]{MilOwe01}
Miller~N.~A., Owen~F.~N., 2001, ApJS, 134, 355
\bibitem[Miller {\rm et~al.}<2002>]{Miletal02}
Miller~N.~A., Ledlow~M.~J., Owen~F.~N., Hill~J.~M., 2002, AJ, 123, 3018
\bibitem[Munoz {\rm et~al.}<2003>]{Munetal03}
Munoz~J.~A., Falco~E.~E., Kochanek~C.~S., Lehar~J., Mediavilla~E., 2003, ApJ,
  594, 684
\bibitem[Murgia {\rm et~al.}<1999>]{Muretal99}
Murgia~M., Fanti~C., Fanti~R., Gregorini~L., Klein~U., Mack~K.-H., Vigotti~M.,
  1999, A\&A, 345, 769
\bibitem[Norris {\rm et~al.}<1985>]{Noretal85}
Norris~R.~P., Baan~W.~A., Haschick~A.~D., Diamond~P.~J., Booth~R.~S., 1985,
  MNRAS, 213, 823
\bibitem[Norris<1988>]{Nor88}
Norris~R.~P., 1988, MNRAS, 230, 345
\bibitem[O'Dea {\rm et~al.}<1991>]{ODeBauSta91}
O'Dea~C.~P., Baum~S.~A., Stanghellini~C., 1991, ApJ, 380, 66
\bibitem[O'Dea<1998>]{ODe98}
O'Dea~C., 1998, PASP, 110, 493
\bibitem[Ojha {\rm et~al.}<2004>]{Ojhetal04}
Ojha~R., Fey~A.~L., Johnston~K.~J., Jauncey~D.~L., Tzioumis~A.~K.,
  Reynolds~J.~E., 2004, AJ, 127, 1977
\bibitem[Orienti {\rm et~al.}<2004>]{Orietal04}
Orienti~M., Dallacasa~D., Fanti~C., Fanti~R., Tinti~S., Stanghellini~C., 2004,
  A\&A, 426, 463
\bibitem[Owsianik {\rm et~al.}<1998>]{OwsConPol98}
Owsianik~I., Conway~J.~E., Polatidis~A.~G., 1998, A\&A, 336, L37
\bibitem[Patnaik {\rm et~al.}<1992a>]{Patetal92a}
Patnaik~A.~R., Browne~I. W.~A., Wilkinson~P.~N., Wrobel~J.~M., 1992a, MNRAS,
  254, 655
\bibitem[Patnaik {\rm et~al.}<1992b>]{Patetal92b}
Patnaik~A.~R., Browne~I. W.~A., Walsh~D., Chaffee~F.~H., Foltz~C.~B., 1992b,
  MNRAS, 259, P1
\bibitem[Peacock \& Wall<1982>]{PeaWal82}
Peacock~J.~A., Wall~J.~V., 1982, MNRAS, 198, 843
\bibitem[Peck \& Taylor<2000>]{PecTay00}
Peck~A.~B., Taylor~G.~B., 2000, ApJ, 534, 90
\bibitem[Peck {\rm et~al.}<2000>]{Pecetal00}
Peck~A.~B., Taylor~G.~B., Fassnacht~C.~D., Readhead~A. C.~S., Vermeulen~R.~C.,
  2000, ApJ, 534, 104
\bibitem[Perlman {\rm et~al.}<1994>]{Peretal94}
Perlman~E.~S., Stocke~J.~T., Shaffer~D.~B., Carilli~C.~L., Ma~C., 1994, ApJ,
  424, L69
\bibitem[Perlman {\rm et~al.}<1996>]{Peretal96}
Perlman~E.~S., Carilli~C.~L., Stocke~J.~T., Conway~J., 1996, AJ, 111, 1839
\bibitem[Perlman {\rm et~al.}<1998>]{Peretal98}
Perlman~E.~S., Padovani~P., Giommi~P., Sambruna~R., Jones~L.~R., Tzioumis~A.,
  Reynolds~J., 1998, AJ, 115, 1253
\bibitem[Perlman {\rm et~al.}<2001>]{Peretal01}
Perlman~E.~S., Stocke~J.~T., Conway~J., Reynolds~C., 2001, AJ, 122, 536
\bibitem[Perucho \& Mart\'{\i}<2002>]{PerMar02}
Perucho~M., Mart\'{\i}~J.~M., 2002, ApJ, 568, 639
\bibitem[Phillips \& Mutel<1982>]{PhiMut82}
Phillips~R.~B., Mutel~R.~L., 1982, A\&A, 106, 21
\bibitem[Polatidis \& Conway<2003>]{PolCon03}
Polatidis~A.~G., Conway~J.~E., 2003, PASA, 20, 69
\bibitem[Polatidis {\rm et~al.}<1995>]{Poletal95}
Polatidis~A.~G., Wilkinson~P.~N., Xu~W., Readhead~A. C.~S., Pearson~T.~J.,
  Taylor~G.~B., Vermeulen~R.~C., 1995, ApJS, 98, 1
\bibitem[Polatidis {\rm et~al.}<1999>]{Poletal99}
Polatidis~A.~G., Wilkinson~P.~N., Xu~W., Readhead~A. C.~S., Pearson~T.~J.,
  Taylor~G.~B., Vermeulen~R.~C., 1999, New Astron. Rev., 43, 657
\bibitem[Puchnarewicz {\rm et~al.}<1992>]{Pucetal92}
Puchnarewicz~E.~M., Mason~K.~O., Cordova~F.~A., Kartje~J.,
  Brabduardi-Raymont~G., Mittaz~J. P.~D., Murdin~P.~G., Allington-Smith~J.,
  1992, MNRAS, 256, 589
\bibitem[Readhead {\rm et~al.}<1996a>]{Reaetal96a}
Readhead~A. C.~S., Taylor~G.~B., Xu~W., Pearson~T.~J., Wilkinson~P.~N., 1996a,
  ApJ, 460, 612
\bibitem[Readhead {\rm et~al.}<1996b>]{Reaetal96b}
Readhead~A. C.~S., Taylor~G.~B., Pearson~T.~J., Wilkinson~P.~N., 1996b, ApJ,
  460, 634
\bibitem[Reid {\rm et~al.}<1995>]{Reietal95}
Reid~A., Shone~D.~L., Akujor~C.~E., Browne~I. W.~A., Murphy~D.~W., Pedelty~J.,
  Rudnick~L., Walsh~D., 1995, A\&AS, 110, 213
\bibitem[Rossetti {\rm et~al.}<2005>]{Rosetal05}
Rossetti~A., Mantovani~F., Dallacasa~D., Fanti~C., Fanti~R., 2005, A\&A, 434,
  449
\bibitem[Rovilos {\rm et~al.}<2003>]{Rovetal03}
Rovilos~E., Diamond~P.~J., Lonsdale~C.~J., Smith~H.~E., 2003, MNRAS, 342, 373
\bibitem[Saikia {\rm et~al.}<1990>]{Saietal90}
Saikia~D.~J., Junor~W., Cornwell~T.~J., Muxlow~T. W.~B., Shastri~P., 1990,
  MNRAS, 245, 408
\bibitem[Saikia {\rm et~al.}<2003>]{Saietal03}
Saikia~D.~J., Jeyakumar~S., Mantovani~F., Salter~C.~J., Spencer~R.~E.,
  Thomasson~P., Wiita~P.~J., 2003, PASA, 20, 50
\bibitem[Sanghera {\rm et~al.}<1995>]{Sanetal95}
Sanghera~H.~S., Saikia~D.~J., Luedke~E., Spencer~R.~E., Foulsham~P.~A.,
  Akujor~C.~E., Tzioumis~A.~K., 1995, A\&A, 295, 629
\bibitem[Sargent<1973>]{Sar73}
Sargent~W. L.~W., 1973, ApJ, 182, L13
\bibitem[Scoville {\rm et~al.}<1998>]{Scoetal98}
Scoville~N.~Z. {\rm et~al.}, 1998, ApJ, 492, L107
\bibitem[Shaya {\rm et~al.}<1994>]{Shaetal94}
Shaya~E.~J., Dowling~D.~M., Currie~D.~G., Faber~S.~M., Groth~E.~J., 1994, AJ,
  107, 1675
\bibitem[Snellen {\rm et~al.}<1999>]{Sneetal99}
Snellen~I. A.~G., Schilizzi~R.~T., Bremer~M.~N., Miley~G.~K., de~Bruyn~A.~G.,
  Rottgering~H.~J., 1999, MNRAS, 307, 149
\bibitem[Snellen {\rm et~al.}<2000>]{SneSchvLa00a}
Snellen~I. A.~G., Schilizzi~R.~T., van Langevelde~H.~J., 2000, MNRAS, 319, 429
\bibitem[Soifer {\rm et~al.}<1984>]{Soietal84}
Soifer~B.~T. {\rm et~al.}, 1984, ApJ, 283, L1
\bibitem[Sowards-Emmerd {\rm et~al.}<2003>]{SowRomMic03}
Sowards-Emmerd~D., Romani~R.~W., Michelson~P.~F., 2003, ApJ, 590, 109
\bibitem[Spencer {\rm et~al.}<1989>]{Speetal89}
Spencer~R.~E., McDowell~J.~C., Charlesworth~M., Fanti~C., Parma~P.,
  Peacock~J.~A., 1989, MNRAS, 240, 657
\bibitem[Stanghellini {\rm et~al.}<1997a>]{Staetal97}
Stanghellini~C., O'Dea~C.~P., Baum~S.~A., Dallacasa~D., Fanti~R., Fanti~C.,
  1997a, A\&A, 325, 943
\bibitem[Stanghellini {\rm et~al.}<1997b>]{Staetal97b}
Stanghellini~C., Bondi~M., Dallacasa~D., O'Dea~C.~P., Baum~S.~A., Fanti~R.,
  Fanti~C., 1997b, A\&A, 318, 376
\bibitem[Stanghellini {\rm et~al.}<1998>]{Staetal98}
Stanghellini~C., O'Dea~C.~P., Dallacasa~D., Baum~S.~A., Fanti~R., Fanti~C.,
  1998, A\&AS, 131, 303
\bibitem[Stanghellini {\rm et~al.}<1999>]{StaODeMur99}
Stanghellini~C., O'Dea~C.~P., Murphy~D.~W., 1999, A\&AS, 134, 309
\bibitem[Stanghellini {\rm et~al.}<2001>]{Staetal01}
Stanghellini~C., Dallacasa~D., O'Dea~C.~P., Baum~S.~A., Fanti~R., Fanti~C.,
  2001, A\&A, 377, 377
\bibitem[Stickel \& Kuhr<1993>]{StiKuh93a}
Stickel~M., Kuhr~H., 1993, A\&AS, 101, 521
\bibitem[Stickel \& Kuhr<1996>]{StiKuh96}
Stickel~M., Kuhr~H., 1996, A\&AS, 115, 1
\bibitem[Stickel {\rm et~al.}<1996>]{Stietal96}
Stickel~M., Rieke~G.~H., Kuhr~H., Rieke~M.~J., 1996, ApJ, 468, 556
\bibitem[Sykes<1997>]{Syk97}
Sykes~C.~M., 1997, Ph.D. Thesis, University of Manchester, UK
\bibitem[Tadhunter {\rm et~al.}<1993>]{Tadetal93}
Tadhunter~C.~N., Morganti~R., di~Serego-Alighieri~S., Fosbury~R. A.~E.,
  Danziger~I.~J., 1993, MNRAS, 263, 999
\bibitem[Taylor \& Peck<2003>]{TayPec03}
Taylor~G.~B., Peck~A.~B., 2003, ApJ, 597, 157
\bibitem[Taylor \& Vermeulen<1997>]{TayVer97}
Taylor~G.~B., Vermeulen~R.~C., 1997, ApJ, 485, L9
\bibitem[Taylor {\rm et~al.}<1994>]{Tayetal94}
Taylor~G.~B., Vermeulen~R.~C., Pearson~T.~J., Readhead~A. C.~S.,
  Henstock~D.~R., Browne~I. W.~A., Wilkinson~P.~N., 1994, ApJS, 95, 345
\bibitem[Taylor {\rm et~al.}<1996a>]{TayReaPea96}
Taylor~G.~B., Readhead~A. C.~S., Pearson~T.~J., 1996a, ApJ, 463, 95
\bibitem[Taylor {\rm et~al.}<1996b>]{Tayetal96}
Taylor~G.~B., Vermeulen~R.~C., Readhead~A. C.~S., Pearson~T.~J.,
  Henstock~D.~R., Wilkinson~P.~N., 1996b, ApJS, 107, 37
\bibitem[Taylor {\rm et~al.}<1997>]{Tayetal96b}
Taylor~G.~B., Vermeulen~R.~C., Readhead~A. C.~S., Pearson~T.~J.,
  Henstock~D.~R., Wilkinson~P., 1997, in Snellen~I. A.~G., Schilizzi~R.~T.,
  Rottgering~H. J.~A., Bremer~M.~N., eds, The Second Workshop on Gigahertz
  Peaked Spectrum and Compact Steep Spectrum Radio Sources.
\newblock Leiden Observatory
\bibitem[Taylor {\rm et~al.}<1998>]{TayWroVer98}
Taylor~G.~B., Wrobel~J.~M., Vermeulen~R.~C., 1998, ApJ, 498, 619
\bibitem[Taylor {\rm et~al.}<2000>]{Tayetal00}
Taylor~G.~B., Marr~J.~M., Pearson~T.~J., Readhead~A. C.~S., 2000, ApJ, 541, 112
\bibitem[Taylor {\rm et~al.}<2005>]{Tayetal05}
Taylor~G.~B. {\rm et~al.}, 2005, ApJS, 159, 27
\bibitem[Thompson {\rm et~al.}<1990>]{ThoDjoCar90}
Thompson~D., Djorgovski~S., de~Carvalho~R., 1990, PASP, 102, 1235
\bibitem[Torniainen {\rm et~al.}<2005>]{Toretal05}
Torniainen~I., Tornikoski~M., Terasranta~H., Aller~M.~F., Aller~H.~D., 2005,
  A\&A, 435, 839
\bibitem[Tornikoski {\rm et~al.}<2001>]{Toretal01}
Tornikoski~M., Jussila~I., Johansson~P., Lainela~M., Valtaoja~E., 2001, AJ,
  121, 1306
\bibitem[Tschager {\rm et~al.}<2000>]{Tscetal00}
Tschager~W., Schilizzi~R.~T., Rottgering~H. J.~A., Snellen~I. A.~G.,
  Miley~G.~K., 2000, A\&A, 360, 887
\bibitem[Tschager {\rm et~al.}<2003>]{Tscetal03}
Tschager~W., Schilizzi~R.~T., Rottgering~H. J.~A., Snellen~I. A.~G.,
  Miley~G.~K., Perley~R.~A., 2003, A\&A, 402, 171
\bibitem[Tzioumis {\rm et~al.}<1989>]{Tzietal89}
Tzioumis~A. {\rm et~al.}, 1989, AJ, 98, 36
\bibitem[Tzioumis {\rm et~al.}<2002>]{Tzietal02}
Tzioumis~A.~K. {\rm et~al.}, 2002, A\&A, 392, 841
\bibitem[Unger {\rm et~al.}<1984>]{Ungetal84}
Unger~S.~W., Pedlar~A., Neff~S.~G., de~Bruyn~A.~G., 1984, MNRAS, 209, 15P
\bibitem[Unger {\rm et~al.}<1986>]{Ungetal86}
Unger~S.~W., Pedlar~A., Booler~R.~V., Harrison~B.~A., 1986, MNRAS, 219, 387
\bibitem[Vermeulen \& Taylor<1995>]{VerTay95}
Vermeulen~R.~C., Taylor~G.~B., 1995, AJ, 109, 1983
\bibitem[Vermeulen {\rm et~al.}<1996>]{Veretal96}
Vermeulen~R.~C., Taylor~G.~B., Redhead~A. C.~S., Browne~I. W.~A., 1996, AJ,
  111, 1013
\bibitem[Wegner {\rm et~al.}<1999>]{Wegetal99}
Wegner~G., Colless~M., Saglia~R.~P., McMahan~R.~K., Davies~R.~L., Burstein~D.,
  Baggley~G., 1999, MNRAS, 305, 259
\bibitem[White \& Becker<1992>]{WhiBec92}
White~R.~L., Becker~R.~H., 1992, ApJS, 79, 331
\bibitem[White {\rm et~al.}<1993>]{WhiKinBec93}
White~R.~L., Kinney~A.~L., Becker~R.~H., 1993, ApJ, 407, 456
\bibitem[Wiklind \& Combes<1997>]{WikCom97}
Wiklind~T., Combes~F., 1997, A\&A, 328, 48
\bibitem[Wilkinson {\rm et~al.}<1994>]{Wiletal94}
Wilkinson~P.~N., Polatidis~A., Readhead~A. C.~S., Xu~W., Person~T.~J., 1994,
  ApJ, 432, L87
\bibitem[Wilkinson<1995>]{Wil95}
Wilkinson~P.~N., 1995, Proc. Natl. Acad. Sci, 92, 11342
\bibitem[Willott {\rm et~al.}<1998>]{Wiletal98}
Willott~C.~J., Rawlings~A., Blundell~K.~M., Lacy~M., 1998, MNRAS, 300, 625
\bibitem[Wills \& Wills<1976>]{WilWil76}
Wills~D., Wills~B.~J., 1976, ApJS, 31, 143
\bibitem[Xu {\rm et~al.}<1994>]{Xuetal94}
Xu~W., Lawrence~C.~R., Readhead~A. C.~S., Pearson~T.~J., 1994, AJ, 108, 395
\bibitem[Xu {\rm et~al.}<1995>]{Xuetal95}
Xu~W., Readhead~A. C.~S., Pearson~T.~J., Polatidis~A.~G., Wilkinson~P.~N.,
  1995, ApJS, 99, 297
\bibitem[Yee {\rm et~al.}<1996>]{Yeeetal96}
Yee~H. K.~C., Ellingson~E., Abraham~R.~G., Gravel~P., Carlberg~R.~G.,
  Smecker-Hane~T.~A., Schade~D., Rigler~M., 1996, ApJS, 102, 289
\bibitem[Zensus {\rm et~al.}<2002>]{Zenetal02}
Zensus~J.~A., Ros~E., Kellermann~K.~I., Cohen~M.~H., Vermeulen~R.~C.,
  Kadler~M., 2002, AJ, 124, 662

\end{thebibliography}

\appendix

\section{Comments on previous related papers}

In what follows, in two separate sections, we discuss the implications of the new parent sample (1743 sources vs. the old 1665) and main sample (157 sources vs. the old 55) sizes and membership to the two main related papers, \scite{papI} and \scite{AugWil01}.

\subsection{Augusto et al.\ (1998)}

Well after publishing this paper, we found that one of the 55-sample sources (B1947+677) had a significantly wrong position.
It should have read (J2000.0): 
19:47:36.2599 (RA) and	67:50:16.928 (dec). This has	been corrected in Tables~4 and~5. In addition, there are now new $\alpha_{1.40}^{4.85}$ values for seven of the \scite{papI} Table~2 (Column (7)) 55-sample sources while five others (B0352+825, B0819+082, B0905+420, B1801+036, B2101+664) should be left with blanks (rather than fill them with spectral indices from sources other than \scite{WhiBec92} and \scite{GreCon91}):
 $-0.02$ (B0218+357); $0.25$ (B0821+394);  $0.24$ (B0831+557);  $0.22$ (B0916+718);  $0.47$ (B1143+446); $0.38$ (B1947+677);  $0.04$ (B2201+044).

As regards the statistical conclusions of \scite{papI} we must revise them by comparing the ``old'' and the ``new'' situations, now that we have both revised the parent sample and the 55-source sample (which, now, has all its sources included in the larger 157-source sample). We focus on the $\alpha_{1.40}^{4.85}$ distribution only. Starting with the parent samples we first note that, while for the new sample we used the maximum number possible of values (1311) for the old one a representative sub-sample of 373 sources was selected and it is from this one that the statistics of \scite{papI} are worked out. Comparing both through a
 KS-test we cannot reject the hypothesis that they are similar. 
Finally, comparing the $\alpha_{1.40}^{4.85}$ distributions for the 157-source (actually  123 values) and 55-source samples through a
 KS-test we cannot reject the hypothesis that they are similar.

\subsection{Augusto \& Wilkinson (2001)}

The only result from  \scite{AugWil01} that is affected by the change from the old 1665-source parent sample to the new 1743-source one relates to their different sizes: the quoted maximum multiple imaging lensing rate on $10^{9.5}$--$10^{10.9} M_{\odot}$ (160--300 mas angular separation of images) of 1:555 (95\% confidence level) actually improves to 1:581 at the same confidence level. As regards to the main result of the paper (limits on the density of compact objects within the above mass range of $\Omega_{CO}<0.1$ at 95\% confidence), this remains unchanged since the 5\% increase in sample size does not cause significant effects.


\end{document}


\bibliographystyle{mnras}

\begin{table}
\label{review}
\caption[this is just to avoid the buffer problem...]{\small The collection of confirmed CSOs from the literature (selected among all sources that have ever been called CSO, MSO, GPS, CSS or CD). Description for each column:
{\bf (1):} Source name; $^{*}$there are comments in the text; {\bf (2), (3):}  source position from the NASA Extragalactic Database (NED); {\bf (4):} the optical host type, when identified; G --- galaxy, Q --- QSO; {\bf (5), (6):}  the source redshift (`p' indicates a photometric estimate) and reference (listed at the end of this caption);    {\bf (7):} classification --- only final for cases with values in column (17); `CSO?' indicates that this source still might be an MSO, depending on its linear size; {\bf (8):} references with the relevant maps that allow final classification (listed at the end of this caption); {\bf (9):} the interferometer(s) used that allow final classification: V (VLA), M (MERLIN) or VL (VLBI); the interferometer that produced the map actually used in the determinations of Columns~(13)--(16) is indicated in bold;   {\bf (10):} the measured largest angular size between either 6$\sigma$ or $5\sigma$ (marked *) contours (5~GHz; exceptions --- other frequencies --- are indicated as superscripts);  {\bf (11):} spectral index $\alpha_{1.40}^{4.85}$ with flux densities (at 1.40~GHz and 4.85~GHz) from White \& Becker (1992) and Gregory \& Condon (1991), respectively;  
{\bf (12):} when possible ($>2$ points), the fitted spectral index to the thin part of the overall spectrum (as in NED); A --- peakless, complex spectrum (v-shaped); in superscript is the estimated spectrum peak (in GHz) while as subscripts we use: $s$ --- steepening spectrum from peak; $v$ --- variable source; 3 --- only three points used; 
  {\bf (13):}  S$^{peak}_{nucleus}$/S$^{peak}_{bright-lobe}$ (5 GHz); exceptions (other frequencies) are indicated as superscripts; {\bf (14):} S$^{peak}_{bright-lobe}$/S$^{peak}_{faint-lobe}$ (5 GHz); exceptions (other frequencies) are indicated as superscripts;  {\bf (15):} arm length ratio (5 GHz);  exceptions (other frequencies) are indicated as superscripts; {\bf (16):} the `arm angle' (5~GHz), measured between  the opposed lobes; exceptions (other frequencies) are indicated as superscripts; {\bf (17):}  the projected linear size in kpc (from Columns~(5) and (10)); 
{\bf (18):}  the log$_{10}$ 1.4~GHz power in W/Hz (data from White \& Becker 1992).
The references for Columns (6) and (8):
 1-- Phillips \& Mutel (1982); 2-- Herbig \& Readhead (1992); 3-- Dallacasa et al.\ (1995); 4-- Fey et al.\ (1996); 5-- Kellerman et al.\ (1998); 6-- Beasley et al.\ (2002); 7-- The NASA Extragalactic Database (NED); 8-- Gallimore et al.\ (1999); 9-- Stanghellini et al.\ (2001); 10-- Stanghellini et al.\ (1997b); 11-- Cotton et al.\ (1995); 12-- Zenzus et al.\ (2002); 13-- Fomalont et al.\ (2000); 14-- Taylor et al.\ (1994); 15-- Dallacasa et al.\ (2002a); 16-- Dallacasa et al.\ (2002b); 17-- Carilli et al.\ (1998); 18-- Owsianik et al.\ (1998); 19-- Taylor et al.\ (2000); 20-- Peck \& Taylor (2000); 21-- Giovannini et al.\ (2001); 22-- Marcha et al.\ (1996); 23-- Sanghera et al.\ (1995); 24-- Hewitt \& Burbidge (1991); 25-- Parkes Catalogue (1990), Australia Telescope National Facility, Wright \& Otrupcek, (Eds); 26-- Xu et al.\ (1995); 27-- Polatidis et al.\ (1995); 28-- Taylor et al.\ (1996a); 29-- Taylor \& Peck (2003); 30-- Snellen et al.\ (2000a); 31-- Stanghellini et al.\ (1997a); 32-- Stickel \& Kuhr (1993); 33-- Stanghellini et al.\ (1999); 34-- Akujor et al.\ (1996); 35-- Conway et al.\ (1992); 36-- Readhead et al.\ (1996a); 37-- Aller et al.\ (1992); 38--- de Vries et al.\ (1995); 39-- Bondi et al.\ (1998); 40-- Augusto et al.\ (1998); 41-- Unger et al.\ (1984); 42-- Unger et al.\ (1986); 
43-- Vermeulen et al.\ (1996); 44-- Wiklind \& Combes (1997); 45-- Perlman et al.\ (1994); 46-- Baum et al.\ (1990); 47-- 
Perlman et al.\ (1996); 48-- de Vries et al.\ (2000); 49-- Snellen et al.\ (1999); 50-- Tadhunter et al.\ (1993); 51-- Tzioumis et al.\ (2002); 52-- Ojha et al.\ (2004); 53-- Tschager et al.\ (2000); 54-- Bartel et al.\ (1984); 55-- Conway et al.\ (1994); 56-- Taylor \& Vermeulen (1997); 57-- Sykes (1997); 58-- Wilkinson (1995); 59-- Orienti et al. (2004); 60-- Gugliucci et al. (2005); 61-- Peck et al. (2000); 62-- Miller et al. (2002); 63-- Giroletti, Giovannini, and Taylor (2005).}
\clearpage
\begin{center}
\begin{scriptsize}
\begin{tabular}{lccccccccccccccccc} \hline \hline
\multicolumn{1}{c}{\bf (1)} & {\bf (2)} & {\bf (3)} & {\bf (4)} & {\bf (5)} & {\bf (6)} & {\bf (7)} & {\bf (8)} & {\bf (9)} & {\bf (10)} & {\bf (11)} & {\bf (12)} & {\bf (13)} & {\bf (14)} & {\bf (15)} & {\bf (16)} & {\bf (17)} & {\bf (18)}  \\
\multicolumn{1}{c}{Name} & RA (J2000.0) & Dec (J2000.0) & Id. & z & Ref. & Class. & Ref. & Tel. & LAS (\arcsec) & $\alpha_{1.40}^{4.85}$  & $\alpha_{thin}$ & S$_n$/S$_{bl}$ & S$_{bl}$/S$_{fl}$ & R & $\phi$&  l & L$_{1.4}$ \\ \hline
4C+40.52		&00h00m53.0815s &+40$\degr$54'01.793'' & 	G	& & & CSO &	6,15,20,59& VL & 0.10  & 0.51& $0.6^{0.4}_s$& 0.19 & 28 & 1.2 & 166  & $<1$ & \\
B0001+478 	& 00h03m46.0412s & +48$\degr$07'04.133'' &  &  &  & CSO & 20 & VL & 0.015 & 0.60& $0.6_3$& 0.08& 1.5& 1.1 & 169$^{\circ}$ & $<1$ & \\
B0026+346  &00h29m14.2424s & +34$\degr$56'32.247'' 	&  G	& 0.517 & 12 &	CSO & 5,6,13,14& VL & 0.040  & 0.23& $0.3^{1}$& 0.040& 3.3& 1.5& 171$^{\circ}$ & 0.21 & 25.9 \\
B0046+316$^*$  & 00h48m47.1438s & +31$\degr$57'25.094'' & G   &  0.015 & 42& CSO &   40,41& M & 0.38           & 0.05& $0_v$& 5.0&2.7 &1.3 & 175$^{\circ}$& 0.11 & 23.0\\
B0108+388   	& 01h11m37.3169s &+39$\degr$06'28.104" & G &  0.668 & 17 & 	CSO	& 12,18,19,46& VL & 0.008$^{15}$  &$-0.94$ & $1.0^{5}_s$&  0.08$^{15}$ & 1.4$^{15}$ & 1.4$^{15}$ & 171$^{\circ}$$^{15}$ & 0.047 & 25.9\\
4C+31.04 & 01h19m34.9991s &+32$\degr$10'50.013" & G&		0.060 & 22 &	CSO	& 11,21,40& VL & 0.11 & 0.47& $0.4^{0.4}$& 0.7& 3.5& 1.1& 173$^{\circ}$& 0.12 & 25.2\\	
B0147+400 & 01h50m19.623s &+40$\degr$17'29.98" & & &  &	CSO	& 15,59& VL & 0.075 & 0.67& $0.6_s$& 0.17& 12& 4.0& 180$^{\circ}$& $<1$ &  \\	
B0201+088    & 02h04m34.7591s & +09$\degr$03'49.259" &  & & &CSO &	6,20 & VL & 0.037 	& 0.75& ---$^{1}$& 1.5&1.1 &1.8 &165$^{\circ}$ & $<1$ & \\
B0222+36    & 02h25m27.351s & +37$\degr$10'27.56" & G & 0.033 & 62 &CSO &	63 & {\bf V}+VL & 1.2$^{8.4}$ 	& 0.53&$0.4_3$&  6.6$^{8.4}$ &3.1$^{8.4}$  &1.1$^{8.4}$  &151$^{\circ}$$^{8.4}$ & 0.74 & 23.7 \\
B0233+434       & 02h37m01.2149s & +43$\degr$42'04.191'' & & & & CSO &  40& M+{\bf VL} & 0.14 & 0.45&$0.5^{0.4}_3$& $<0.006$ & 2.2& ---  & ---&  $<1$ & \\
B0352+825        & 04h02m12.6736s & +82$\degr$41'35.103'' &    & &   &    CSO	&  40& M+{\bf VL} & 0.069 & ---&---& 5.4&3.6 &1.8 &180$^{\circ}$ & $<1$ & \\
B0423$-$163          & 04h25m53.5726s & $-16\degr$12'40.248" & & & & CSO	&29& VL & 0.096  & ---& $0.6_s$ & 0.035& 7.1& 2.7& 162$^{\circ}$& $<1$ & \\
B0424+414$^*$          & 04h27m46.0455s &+41$\degr$33'01.099" &	& & & CSO &	20& VL & 0.006$^{15}$ & $-0.10$& ---$^{5}$& 39$^{15}$ & 1.4$^{15}$ & 1.5$^{15}$ & 162$^{\circ}$$^{15}$ & $<1$ &  \\
B0428+205          & 04h31m03.7585s	&+20$\degr$37'34.189"&	G&0.219 &25 & CSO &	3,58& {\bf M}+VL & 0.32$^{1.7}$ *& $0.20$& $0.6^{1}_s$&  0.06$^{1.7}$ & 31$^{1.7}$ & 4.6$^{1.7}$ & 177$^{\circ}$$^{1.7}$ & 0.99 & 26.3\\
B0500+019$^*$   & 05h03m21.1971s &+02$\degr$03'04.677" &	G	&0.585& 17&	CSO & 9,31& VL & 0.017 &0.09 &$0.5^{3}$&  1.7&2.2 & 4.1&	139$^{\circ}$& 0.094 & 26.5 \\
B0646+600$^*$           & 06h50m31.2543s &+60$\degr$01'44.555"& 	Q&	0.455& 32&	CSO& 26,33,34& VL & 0.009& $-0.59$&$0.5^{4}_s$&  0.83& 113& 1.1& 158$^{\circ}$& 0.044 & 25.7	\\
B0703+468$^*$            &07h06m48.037s &+46$\degr$47'56.22"& Q? & &	& CSO& 15,23,59& V+{\bf VL} & 0.065 	& 0.67& $0.5^{0.4}_{3,s}$&  $<0.005$ & 1.6&--- & ---& $<1$ & \\
B0710+439$^*$             & 07h13m38.1641s &+43$\degr$49'17.205" &	G	&0.518& 35&	CSO& 19,28,35,36& VL & 0.026 & 0.09& $0.4^{2}$& 0.68& 4.3& 1.9&176$^{\circ}$ & 0.14 & 26.4 \\
4C+39.17 	&07h25m49.993s &+39$\degr$17'25.47" & G& & &	CSO? &	16& V+{\bf VL} & 0.42$^{1.7}$ &1.26 &$1.1^{0.4}_{3,s}$&  $<0.02^{1.7}$&20$^{1.7}$ &--- &--- & ? & \\
B0750+535  & 07h54m15.2176s &+53$\degr$24'56.449" &	 &  &  & CSO &  20	& VL & 0.026 & 0.59&$0.8^{1}_{3,s}$&  0.18& 1.0&2.3 &174$^{\circ}$ & $<1$ & \\
 B0840+424A	& 08h43m31.6374s &+42$\degr$15'29.525"  & & & &	CSO &  15,59& VL & 0.12  	& 0.64& $0.6^{0.4}_s$& $<0.2$&19 &--- &--- & $<1$ & \\
B1031+567   & 10h35m07.0403s &+56$\degr$28'46.798" &	G&	0.45& 37&	CSO & 13,19,30& VL & 0.038 &0.31 &$0.5^{0.8}$&  $<0.004$&3.1 &--- & ---& 0.19 & 26.3 \\
J1111+1955	& 11h11m20.0658s &+19$\degr$55'36.001" & G&0.299 &67 &	CSO &	20&  VL & 0.020$^{8.4}$ &0.46 &$0.5^{1}_3$&  $<0.02^{8.4}$&2.0$^{8.4}$ &--- &--- & 0.077 & 25.9 \\
4C+14.41    & 11h20m27.8043s &+14$\degr$20'55.020" &	G	& 0.362 & 38&	CSO & 6,23,39& {\bf M}+VL & 0.11$^{23}$  & 0.82&$0.5^{0.4}_s$&  1.1$^{23}$ & 2.9$^{23}$ & 1.2$^{23}$ & 165$^{\circ}$$^{23}$ & 0.48 &26.4 \\
B1208$-$192 & 12h11m31.801s &$-19\degr34'19.95"$ &		Q & 2.054	& 24&	CSO	&   31& VL & 0.023$^{15}$  & --- & $0.6$& 0.042$^{15}$ & 1.9$^{15}$& 1.2$^{15}$& 180$^{\circ}$$^{15}$& 0.16 & ---\\
4C+32.44   &		13h26m16.5124s & +31$\degr$54'09.514'' 	&  G & 0.370 & 2 & CSO & 1,3,4,5 & VL & 0.071$^{8.6}$  & 0.61 &$0.6^{0.4}$&  $<0.02^{8.6}$& 2.0$^{8.6}$& ---& ---& 0.31 & 27.1\\
4C+62.22 &	14h00m28.593s &+62$\degr$10'40.11''& 	G	&0.431 & 7&		CSO & 3,28& VL & 0.060$^{8.4}$ & 0.70& $0.7^{0.6}$& 1.0$^{8.4}$ & 1.1$^{8.4}$ & 2.5$^{8.4}$ &  171$^{\circ}$$^{8.4}$  &  0.29 &26.6 	\\
Mrk 668  & 14h07m00.3944s & +28$\degr$27'14.690'' & G & 0.077 & 8 &	CSO & 5,9,10& VL & 0.009$^{15}$  & $-0.93$ &$1.1^{6}_v$&  0.013$^{15}$ & 61$^{15}$ &1.2$^{15}$  & 176$^{\circ}$$^{15}$  & 0.012 & 25.9 \\ 
\hline
\end{tabular}
\end{scriptsize}
\end{center}
\end{table}
\clearpage
\begin{table}
\begin{center}
\begin{scriptsize}
\begin{tabular}{lccccccccccccccccc} \hline \hline
\multicolumn{1}{c}{\bf (1)} & {\bf (2)} & {\bf (3)} & {\bf (4)} & {\bf (5)} & {\bf (6)} & {\bf (7)} & {\bf (8)} & {\bf (9)} & {\bf (10)} & {\bf (11)} & {\bf (12)} & {\bf (13)} & {\bf (14)} & {\bf (15)} & {\bf (16)} & {\bf (17)} & {\bf (18)}  \\
\multicolumn{1}{c}{Name} & RA (J2000.0) & Dec (J2000.0) & Id. & z & Ref. & Class. & Ref. & Tel. & LAS (\arcsec) & $\alpha_{1.40}^{4.85}$  & $\alpha_{thin}$ & S$_n$/S$_{bl}$ & S$_{bl}$/S$_{fl}$ & R & $\phi$&  l & L$_{1.4}$ \\ \hline
B1413+135 	& 14h15m58.8174s &+13$\degr$20'23.713''& 	Q&	0.247 & 44& CSO	& 5,13,20,45,47& VL & 0.18$^{1.7}$ & 0.29& A & 1.6$^{1.7}$ & 22$^{1.7}$ & 2.5$^{1.7}$ & 135$^{\circ}$$^{1.7}$ & 0.61 & 25.0 \\
B1543+005	& 15h46m09.5314s &+00$\degr$26'24.613'' &  G &	0.556 & 48& CSO & 6,13,20,60& VL & 0.017 &0.26 &0.4 & 10&1.3 &1.6	&134$^{\circ}$   & 0.092 & 26.0  \\
B1732+094 	& 17h34m58.3769s & +09$\degr$26'58.259'' &  G	& 0.61$^{p}$   &33 & CSO &	6,20,33& VL & 0.017&0.01 &$0.1_3$&  $<0.01$&2.0 &--- &--- & 0.10 & 26.1  	\\
B1801+036 & 18h03m56.2829s & +03$\degr$41'07.575'' & G  & &   &    CSO?	&  40& M & 1.35  & ---&---& 3.8&8.0 &4.1 &155$^{\circ}$ & ? & \\ 
B1819+671 	& 18h19m44.391s &+67$\degr$08'47.18'' & 	G	& 0.220 &49& CSO &	30 & VL & 0.025 &0.83 &---$^{0.6}$&  0.091& 2.0& 1.1& 180$^{\circ}$& 0.078 & 25.9\\
B1824+185 	& 18h26m17.7108s &+18$\degr$31'52.889'' &  &  &  & CSO &  			6,20,60& VL & 0.069 & 0.52& $0.4_3$& 0.043&14 &1.8 & 177$^{\circ}$ & $<1$ & \\
B1843+356  	&18h45m35.1088s& +35$\degr$41'16.726'' &	G	& 0.764 & 43 & CSO & 6,27,28& VL & 0.034$^{1.6}$  & 0.21&$0.4_{3,s}$&  0.069$^{1.6}$ & 64$^{1.6}$ & 1.0$^{1.6}$ &149$^{\circ}$$^{1.6}$  & 0.21 & 25.7\\
B1928+681  & 19h28m20.5502s & +68$\degr$14'59.247'' &   & &      & CSO?	&  40& M+{\bf VL} & 0.17 &0.41 &$0.5^{0.3}$& 6.8 &2.0 &2.3 &168$^{\circ}$ & ? & \\
B1934$-638^*$	& 19h39m25.0261s & $-63\degr$42'45.625'' & 	G	& 0.183 & 50 & CSO &	51,52& VL & 0.075 & ---&$0.9^{1}_s$&  $<0.008$&1.4 &--- &--- &  0.20& --- \\
B1946+708  	&19h45m53.5199s &+70$\degr$55'48.732'' & 	G	& 0.101 & 49& CSO &	30,56& VL & 0.038 & 0.29&$0.4^{1}_3$&  0.25&14 &2.5 & 169$^{\circ}$ & 0.064 & 26.2 \\	
J1947+678	&19h47m36.2599s&	+67$\degr$50'16.928''& & & & CSO? & 40,57& {\bf M}+VL & 0.52	& 0.38& $0.5_3$& 1.1& 18&1.3 & 161$^{\circ}$& ? &  \\		
B2021+614  	& 20h22m06.6817s & +61$\degr$36'58.804'' &  G & 0.227 & 24 & CSO &27,53,54,55& VL & 0.012&$-0.20$ &$0.4^{6}_v$&  0.83& 9.7&2.3 &148$^{\circ}$  & 0.038 & 25.9 	\\
B2352+495  & 23h55m09.4586s & +49$\degr$50'08.341'' & 	G	& 0.237  & 35 & CSO &  13,19,28,36& VL & 0.055  &0.33 &$0.6^{1}_s$& 9.1 &1.4 &1.7 &180$^{\circ}$  &0.18 & 25.5 \\ \hline
\end{tabular}
\end{scriptsize}
\end{center}
\end{table}

\newpage